\documentclass[
  journal=pasa,
  manuscript=research-paper, %% or "review"
  year=2022,
  volume=xx,
]{cup-journal}

\usepackage{microtype,siunitx,booktabs}
\title{SkyHopper mission science case I: Identification of high redshift Gamma-Ray Bursts through space-based near-infrared afterglow observations
}

\author{M. Thomas}
\affiliation{School of Physics, The University of Melbourne, VIC 3010, Australia}
\email[M. Thomas]{thomasm3@student.unimelb.edu.au}

\author{M. Trenti}
\affiliation{School of Physics, The University of Melbourne, VIC 3010, Australia}
\alsoaffiliation{Australian Research Council Centre of Excellence for All-Sky Astrophysics in 3-Dimensions, Australia}

\author{J. Greiner}
\affiliation{Max-Planck-Institut für extraterrestrische Physik, Giessenbachstr. 1, D-85740 Garching, Germany}

\author{M. Skrutskie}
\affiliation{Department of Astronomy, University of Massachusetts, Amherst, MA 01003; Department of Astronomy, P.O. Box 400325, University of Virginia, Charlottesville, VA 22904-4325}

\author{Duncan A. Forbes}
\affiliation{Centre for Astrophysics \& Supercomputing, Swinburne University, Hawthorn VIC 3122, Australia}

\author{S. Klose}
\affiliation{Th{\"u}ringer Landessternwarte Tautenburg, Sternwarte 5, 07778 Tautenburg, Germany}

\author{K. J. Mack}
\affiliation{Physics Department, North Carolina State University, Raleigh, North Carolina, 27695, USA}

\author{R. Mearns}
\affiliation{School of Physics, The University of Melbourne, VIC 3010, Australia}

\author{B. Metha}
\affiliation{School of Physics, The University of Melbourne, VIC 3010, Australia}
\alsoaffiliation{Australian Research Council Centre of Excellence for All-Sky Astrophysics in 3-Dimensions, Australia}

\author{\mbox{G. Tagliaferri}}
\affiliation{INAF -- Osservatorio Astronomico di Brera, Via Bianchi 46, I-23807 Merate, Italy}

\author{N. Tanvir}
\affiliation{School of Physics and Astronomy, University of Leicester, University Road, Leicester, LE1 7RH, UK}

\author{E. Skafidas}
\affiliation{Department of Electrical and Electronic Engineering, The University of Melbourne, VIC 3010, Australia}

\doi{10.1017/pasa.2020.32}

\received {dd Mmm YYYY}
\revised  {dd Mmm YYYY}
\accepted {dd Mmm YYYY}
\published{dd Mmm YYYY}

 %% First letter not capped
\begin{document}

\begin{abstract}
Long-duration gamma-ray burst (GRB) afterglow observations offer cutting-edge opportunities to characterise the star formation history of the Universe back to the epoch of reionisation, and to measure the chemical composition of interstellar and intergalactic gas through absorption spectroscopy. The main barrier to progress is the low efficiency in rapidly and confidently identifying which bursts are high redshift ($z > 5$) candidates before they fade, as this requires low-latency follow-up observations at near-infrared wavelengths (or longer) to determine a reliable photometric redshift estimate. Since no current or planned gamma ray observatories carry near-infrared telescopes on-board, complementary facilities are needed.  
So far this task has been performed by instruments on the ground, but sky visibility and weather constraints limit the number of GRB targets that can be observed and the speed at which follow-up is possible. 
In this work we develop a Monte Carlo simulation framework to investigate an alternative approach based on the use of a rapid-response near-infrared nano-satellite, capable of simultaneous imaging in four bands from $0.8$ to $1.7\mu$m (a mission concept called SkyHopper). 
Using as reference a sample of 88 afterglows observed with the GROND instrument on the MPG/ESO telescope, we find that such a nano-satellite is capable of detecting in the H band (1.6 $\mu$m) $72.5\% \pm 3.1\%$ of GRBs concurrently observable with the Swift satellite via its UVOT instrument (and $44.1\% \pm 12.3\%$ of high redshift ($z>5$) GRBs)  within 60 minutes of the GRB prompt emission. This corresponds to detecting $\sim 55$ GRB afterglows per year, of which 1-3 have $z > 5$.
These rates represent a substantial contribution to the field of high-$z$ GRB science, as only 23 $z > 5$ GRBs have been collectively discovered by the entire astronomical community over the last $\sim 24$ years. Future discoveries are critically needed to take advantage of next generation follow-up spectroscopic facilities such as 30m-class ground telescopes and the James Webb Space Telescope. Furthermore, a systematic space-based follow-up of afterglows in the near-infrared will offer new insight on the population of dusty (`dark') GRBs which are primarily found at cosmic noon ($z\sim 1-3$). 
Additionally, we find that launching a mini-constellation of 3 near-infrared nano-satellites would increase the detection fraction of afterglows to $\sim 83\%$ and substantially reduce the latency in the photometric redshift determination.
\end{abstract}

\section{Introduction}
\label{sec:intro}

    Long-duration gamma-ray bursts (GRBs) are some of the most luminous explosions in the observable universe. These events are relativistic, jetted explosions of very massive stars at the end of their lives\footnote{Long GRBs are physically distinct from both short gamma-ray bursts, which are associated with the merger between two compact objects \citep{Gehrels_2005, Rueda_2018} and the more recent class of `ultra-long' GRBs \citep{Schady_2017}, neither of which are discussed in this paper.} \citep{Woosley_2006} which in principle are detectable out to redshifts of $z = 10-20$ \citep{Lamb_2000} as they outshine their host galaxy by several orders of magnitude at their peak brightness.
    GRB afterglows decay in brightness rapidly and are generally visible only for a few days after the initial burst\footnote{This statement does not apply to the radio wavelengths, where GRB afterglows can be observed for hundreds of days after the initial burst, e.g. see \citet{Rhodes_2020}.} \citep{Gehrels_2009}, but opportunistic afterglow observations offer many unique opportunities to both directly and indirectly study the high redshift universe, provided that rapid spectroscopic follow-up observations are performed.
    
    Because long GRBs originate from the deaths of short-lived massive stars they serve to randomly sample sites of star formation throughout the universe, providing an alternative means to estimate the cosmic star formation rate (SFR), albeit with the possible presence of a metallicity bias in GRB production efficiency \citep{Robertson_2012, Trenti_2013}. Using GRBs to measure the SFR becomes particularly valuable at redshift $z > 5$, where other tracers of the SFR such as Lyman-$\alpha$ emitters and Lyman-break galaxies become scarcer and prone to selection effects \citep{Stanway_2008}.
    
    GRBs can also act as beacons for identifying faint high redshift galaxies with a precise redshift determination from afterglow spectroscopy. Their association with massive star death means that GRBs select star forming host galaxies independently from their host luminosities \citep{Klose_2004}. By performing deep follow-up imaging on GRBs that have both high accuracy position and redshift measurements, it is possible for observers to investigate the luminosity function of high redshift  galaxies \citep{Trenti_2012, Tanvir_2012, McGuire_2016}. Such measurements give insight into the fraction of star formation occurring in very faint galaxies, beyond the sensitivity limit of imaging surveys in blank fields, and have the potential to constrain the faint end of the galaxy luminosity function, which is critical to determine the contribution of such galaxies to cosmic reionisation.
    
    The extreme luminosity of GRB afterglows make them powerful probes of interstellar gas in their host galaxies, which has been used on many occasions to characterise the interstellar medium in distant galaxies (e.g., \citealt{Klose_2004, Berger_2005, Vreeswijk_2007, Fox_2008, Prochaska_2008}).
    %to a level of detail that will be achieved by other means only once $D > 20$m telescopes come online \citep{Gehrels_2009}.
    Furthermore, GRB afterglows have a high intrinsic UV luminosity and exhibit a featureless power-law spectrum in their rest frame UV region \citep{Sari_1998}, making them natural probes of the neutral hydrogen fraction in the IGM through both observations of damping wings in the Lyman-$\alpha$ line and from analysis of the Lyman forest \citep{Miralda_Escude_1998, Mesinger_2008, McQuinn_2008, Hartoog_2015, Lidz_2021}. In this respect GRBs provide an opportunity to investigate the earliest stages of reionisation (a stage which is challenging to probe with quasar spectroscopy) as GRBs have been observed beyond the redshift of the most distant known quasars \citep{Tanvir_2009, Cucchiara_2011, Wang_2021}, and are possibly present at $z>10$ (e.g. from massive Population III stars) at a time when quasars are not expected to be active yet. 
    
    The main challenge facing astronomers wishing to leverage the utility of GRBs during the epoch of reionisation is the observational difficulty of identifying them. Measuring the redshift of a GRB requires afterglow observations at optical/IR wavelengths, since there are no spectral features in the prompt emission phase (the initial high energy photon emission phase of the GRB). Systematically identifying $z > 5$ GRBs through afterglow spectroscopy is challenging since those objects are rare, accounting for $\lesssim$ 6\% of the GRB population \citep{Greiner_2011, Wanderman_2010}, and their typical brightness implies that 6-10m class telescopes are needed, in particular for the most interesting objects at $z\gtrsim 7$ (e.g. \citealt{Tanvir_2009, Salvaterra_2009}, where the Lyman-$\alpha$ line is shifted beyond $1\mu$m). To further complicate the scenario, given the afterglow's power-law decay in brightness any redshift determination must be done rapidly, within hours of the initial burst.
    
    Since its launch in November 2004 the Neil Gehrels Swift satellite \citep{Gehrels_2004} (hereon referred to as Swift) has been the largest contributor to the rapid identification of GRB targets with its ability to detect and instantly localise GRBs in large numbers ($\sim 100$ per year) using its Burst Alert Telescope (BAT) as well as to observe the GRB afterglow using its X-Ray Telescope (XRT) and UV/Optical Telescope (UVOT) within a few hundred seconds of the onset of the prompt emission. However, for $z \gtrsim 5$ sources the Lyman break is redshifted beyond Swift's UVOT wavelength coverage. Furthermore, one cannot infer the presence of a high redshift GRB from a Swift non-detection since such events are observationally degenerate with `Dark' GRBs, which are undetected at optical--NIR wavelengths due to dust extinction along the line of sight and account for $\sim 25-40\%$ of the GRB population (depending on the definition used) \citep{Klose_2000, Greiner_2011, Perley_2013}. Therefore, observations of the GRB afterglow in the near-infrared wavelengths are critically required to establish that a GRB originates at $z \gtrsim 5$. 
    
    The Gamma-Ray Burst Optical/Near-Infrared Detector (GROND) is a 7-channel imager at the  ground-based MPG/ESO 2.2 metre telescope designed with the specific purpose of observing GRB afterglows in the visible and near-infrared, with demonstrated success \citep{Greiner_2008}. However, ground-based observatories suffer several substantial disadvantages in the rapid follow-up of GRB afterglows: they only have access to the fraction of sky visible overhead, they can only make observations at night time, and they depend heavily on the weather. These factors mean that a single ground-based facility - even if optimally designed and located such as GROND - only has the capability to follow-up promptly a small fraction of the GRBs detected by a satellite such as Swift. 
    
    Nano-satellite missions are becoming more common across multiple fields of astronomy, from high precision photometric observations of exoplanet transits in the optical wavelengths (ASTERIA, see \citealt{knapp2020}) to high-energy astrophysics. For the latter, several missions are underway to use nano-satellite missions to detect the prompt X-ray and gamma ray emission from GRBs, e.g., the GRID mission \citep{Wen_2019} which recently made its first detection of GRB210121A \citep{Wang_2021b}, the CAMELOT mission \citep{Werner_2018} which made its first successful GRB detection of GRB210807A with its  GRBAlpha instrument\footnote{\url{https://gcn.gsfc.nasa.gov/gcn3/30624.gcn3}}, the HERMES Technologic and Scientific Pathfinder constellation \citep{Fiore_2020} scheduled to launch 6-satellites in 2023, with one additional nanosatellite (SpIRIT) carrying the HERMES instrument being developed at the University of Melbourne\footnote{https://spirit.research.unimelb.edu.au/}, and the BurstCube mission \citep{Racusin_2017} which is anticipated to launch in 2022. 
    
    In this paper we propose the use of a near-infrared space telescope on a nano-satellite with rapid re-pointing and low-latency communication capabilities to promptly detect GRB afterglows beyond $\sim 1 \mu$m and determine high-$z$ GRB candidates from photometric redshift measurements. Given the high signal-to-noise near-infrared image quality afforded by observing from space (above the atmospheric foreground), a relatively small ($\sim 0.15$m aperture) telescope has the same point source sensitivity in the $H$-band as a $\sim 2$m class telescope at ground level. 
    
    Specifically, we investigate one of the science goals of the SkyHopper mission concept\footnote{\url{https://skyhopper.research.unimelb.edu.au/}}, which aims to systematically follow-up GRBs identified by  \textit{Swift} and future gamma/X-ray satellites (such as the SVOM mission; see  \citealt{Wei_2016}). By combining a GRB detection in the near-infrared with the UV/optical photometry from Swift, a fast photometric redshift estimate of the GRB redshift is possible by determining the location of the Lyman break in one of the observed bands (e.g. \citealt{Kruhler_2011}). In this context, a near-infrared nano-satellite such as SkyHopper would complement the existing and future observational infrastructure to allow for efficient detection of high redshift GRBs.
    
    This paper focuses on modelling the expected performance of rapid-response near-infrared observations from low-Earth orbit, thus quantifying the potential to use a near-infrared nano-satellite to observe GRB afterglows. In Section \ref{sec:Simulating Gamma-Ray Burst Events} we outline the models we use to construct GRB templates for simulated observations. In Section \ref{sec:Satellite Mission Scenario & Telescope Parameters} we define the specific mission scenario and telescope parameters used in this work. In Section \ref{sec:Observational Pipeline} we describe the simulation framework used for line-of-sight access calculations and performing mock afterglow observations. In Section \ref{sec:Optimising GRB Observations} we explore a range of different approaches to identify an optimal exposure strategy to detect GRB afterglows for use on-board a rapid response near-infrared nano-satellite. In Section \ref{sec:Results and Discussion} we discuss the key results and highlights the expected science opportunities that would be enabled by a SkyHopper-like space telescope. The main conclusions from this work are presented in Section \ref{sec:Conclusion}. 
    
\section{Simulating Gamma-Ray Burst Events}
\label{sec:Simulating Gamma-Ray Burst Events}

    % In order to evaluate the effectiveness of GRB afterglow observations with ground and space telescopes, we construct a simplified but effective procedure to generate realistic models of GRB events in the sky. The GRB observables relevant to this work are the redshift of the burst, the near-infrared afterglow luminosity and the early-time near-infrared light curve decay indices. While observations point towards a correlation be- tween afterglow luminosity and light curve decay indices (e.g. Oates et al. 2017), numerous factors make it difficult to precisely quantify this correlation at the time of writing (limited sample size of UV/Optical afterglow observations, assumptions regarding afterglow off axis emissivity modelling, k-correction type effects, etc.) and so for simplicity we treat them independently in this work.
    
    % /////
    
    In order to compare the effectiveness of GRB afterglow observations with ground and space telescopes, we construct a simplified but effective procedure to generate realistic models of GRB events. The GRB observables relevant to this work are the comoving redshift distribution of long GRBs, the near-infrared afterglow luminosity, and the early time near-infrared afterglow light curve decay indices.
    
    Due to the complexities in accurately simulating the detection of high energy GRB photons by the Swift BAT we shall exclude models of GRB prompt emission from our simulation, despite the GRB prompt emission (both peak and isotropic energy release) being positively correlated with optical afterglow luminosity and negatively correlated with afterglow decay rate \citep{Oates_2015}. By neglecting the correlation between prompt and afterglow luminosity we are likely under-estimating the performance of our nano-satellite, as those bursts detected by Swift must meet the minimum prompt emission flux threshold of Swift's BAT and therefore have a higher afterglow luminosity, making them easier to detect. Conversely, by neglecting the anti-correlation to afterglow decay rate we slightly over-estimate the number of detections since brighter bursts decay more rapidly than fainter ones. We leave the inclusion of these correlations to future works, as simulating Swift's detection of high energy photons is beyond the scope of this work.
    
    % Neglecting these correlations means that we may under-estimate the performance of our nano-satellite, especially in the detection of high redshift bursts, since all GRBs must release enough energy in their prompt emission phase to be detected by Swift, which correlates to their afterglows being significantly brighter and therefore easier to detect.
    
    % However, these bright GRBs also decay in brightness more rapidly, making them more difficult to detect the later they are observed. 
    
    We further choose to exclude the correlation between afterglow luminosity and light curve decay indices \citep{Oates_2015}. Numerous factors make it difficult to precisely quantify this correlation at the time of writing (limited sample size of UV/Optical afterglow observations, assumptions regarding afterglow off axis emissivity modelling, k-correction type effects, etc.) and so for simplicity we treat the two observables as independent in our simulation.
    
    % Furthermore, due to the complexities in accurately simulating the detection of high energy GRB photons by the Swift BAT we shall exclude any modelling of GRB prompt emission from our simulation, despite the correlations between prompt emission peak energy ($E_{\text{peak}}$) and isotropic-equivalent energy release ($E_{\text{iso}}$), and optical afterglow luminosity and decay rate \citep{Oates_2015}. Neglecting these correlations may mean that we under-estimate the performance of an infrared nano-satellite, especially in detecting high redshift GRBs, since any burst with a high enough $E_{\text{iso}}$ to exceed Swift's flux limit is likely to have a brighter optical afterglow and be easier to detect with a small aperture telescope. However given bright GRBs also decay more quickly it is difficult to predict the outcome of these correlations on the results of this simulation, and should be left to a future work which incorporates a more detailed model of GRB prompt emission and observation. 
    
    By neglecting GRB prompt emission modelling and assuming that Swift is able to detect every accessible burst, our simulation will unrealistically increase the total number of bursts detected by Swift. We shall therefore present our results as the fraction of Swift triggers successfully detected by the near-infrared nano-satellite instead of the total number of GRB detections.

    \subsection{Redshift Distribution}
    \label{subsec:Redshift Distribution}
    
        To generate a sample GRB population with a realistic redshift distribution we draw on the work of \cite{Wanderman_2010}, who derive the differential co-moving space density of GRBs at redshift $z$:
        \begin{equation}
                R_{GRB}(z) = 
                \begin{cases}
                    (1+z)^{n_1} &
                    z \leq z_1 \\
                    (1+z)^{n_1 - n_2} (1+z)^{n_2} 
                    &z > z_1\\
                \end{cases}
        \end{equation}
        With $z_1 = 3.11$, $n_1 = 2.07$, $n_2 = -1.36$.
        
        The observed distribution of GRB redshifts can then be calculated by multiplying this function by the co-moving volume element $dV/dz$ and correcting for cosmological time dilation:
        
        \begin{equation}\label{eq:GRB Redshift Distribution}
            R(z) = \frac{R_{GRB}(z)}{(1+z)}\frac{dV(z)}{dz}
        \end{equation}
        
        To generate the redshift of each simulated GRB event we randomly sample from the probability distribution described by Equation \ref{eq:GRB Redshift Distribution} over the domain $z \in \{0,10\}$.

        \subsection{Afterglow Light Curve}
        \label{subsec: Afterglow Light Curve}
            
            Modelling the observed near-infrared light curve of a GRB afterglow for a high redshift ($z > 5$) GRB is an equivalent problem to modelling the rest frame UV/optical light curve due to the cosmological redshifting of light. As such, we draw on the work of \cite{Oates_2009}, who analysed the statistical properties of early-time GRB light curves observed by Swift UVOT. The Oates sample consists of 27 long GRB afterglows observed by Swift between 2005-2007 with redshifts ranging from $0.44-4.41$, which corresponds to sampling rest frame wavelengths as low as $\sim 90$nm (any wavelengths shorter than this are absorbed by neutral hydrogen in the host galaxy and IGM). While rest frame wavelengths above $\sim 250$nm are redshifted beyond the $H$-band for bursts with $z > 5$, the sample exhibits no substantial evolution in the light curve decay index as a function of rest frame wavelength (though with the small sample size it is difficult to state this with confidence). We therefore adopt the results of \citealt{Oates_2009} under the assumption that there is no differential colour evolution within the intrinsic UV afterglow spectrum
                %\footnote{We are neglecting evolution of dust absorption and reddening, but given that star forming galaxies at cosmic dawn have a lower dust content of those at cosmic noon \citep{Bouwens_2015}, our assumption is conservative in the context of estimating the detectability of high redshift GRB afterglows.} 
            \citep{Sari_1998, Kumar_2015}, and that the afterglow light curve does not undergo significant evolution with redshift in the emission rest frame.
                
                % Therefore, we take the results of \citealt{Oates_2009} under the assumption that the rest frame afterglow light curve does not undergo significant evolution with redshift, and that there is no differential colour evolution within the intrinsic UV afterglow spectrum \citep{Sari_1998, Kumar_2015}\footnote{We are neglecting evolution of dust absorption and reddening, but given that star forming galaxies at cosmic dawn have a lower dust content of those at cosmic noon \citep{Bouwens_2015}, our assumption is conservative in the context of estimating the detectability of high redshift GRB afterglows.}.
            
            Motivated by the results from \cite{Oates_2009} we model the afterglow in the observer frame using a broken power-law with a break at $t_{\text{obs}} = 500$ sec, separating the light-curve into an `early-time' and a `late-time' decay phase, where the observer-frame afterglow flux $F$ varies as: 
            
            \begin{align}\label{eq:light curve}
                F(t) \:\propto \: t^{-\alpha} \:\:\:\ \text{   ,   } \:\:\:\:\:\: \alpha = \begin{cases}
                    0 , &t_{\text{obs}} < 500 \text{ sec} \\
                    0.85 , &t_{\text{obs}} > 500 \text{ sec}
                \end{cases}
            \end{align}
            
            For the early-time decay curve we have simplified the results of \cite{Oates_2009}, who found that some events decay rapidly and some experience re-brightening during this phase. To take an average of this behaviour we model the decay slope as flat during this epoch because it also serves to impose a flux cutoff on the model, preventing the magnitude of the afterglow from diverging at early times (as it would in a power-law model with $\alpha>0$) and thus potentially artificially inflating the number of GRBs detected by the nano-satellite at very early times. The value of the late-time decay index is set in accordance with the findings of \cite{Oates_2009}.
            
            To test the validity of the modelling assumption that the light curve is flat for $t < 500$ sec, we will also test the case of a rising light curve at early times i.e., $\alpha_{t<500s} = -0.2$ in accordance with the maximum early-time decay index from \citet{Oates_2009}.
            
        \subsection{Afterglow Brightness}
        \label{subsec:Modelling Afterglow Emission}

            Ideally, the observed H-band magnitude of the afterglow would be simulated by sampling from an early-time UV afterglow luminosity function, thereby calculating the observed brightness of the event by taking into account the jet orientation and off-axis emissivity, the extinction along the line of sight and cosmological redshifting.
            
            However, this approach contains many points of uncertainty. Preliminary work has been carried out to derive an afterglow luminosity function (e.g. \citealt{Oates_2009, Wang_2013}), but the results are affected both by limited number of observed GRBs, and by assumptions on the GRB central engine theoretical modelling which affect off-axis emission (e.g. for a purely hydrodynamic jet, see \citealt{Eerten_2010}), an area which is not yet well understood.
            
            Given these uncertainties we instead sample the H-band magnitude of the afterglow directly from a simplified probability distribution and apply a series of corrections to the observed magnitude in order to account for the effects of the distance to the event, k-correction, time dilation and host-galaxy extinction.
            
\begin{figure}[t!]
                \begin{center}
                \includegraphics[width=\columnwidth]{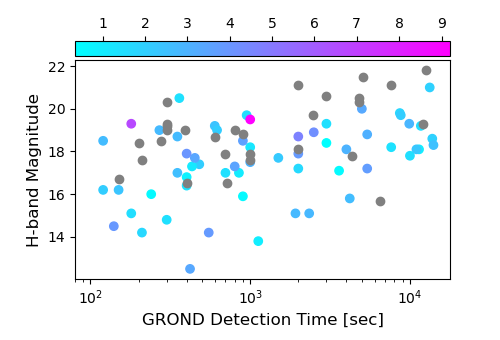}
                \\
                \includegraphics[width = \columnwidth]{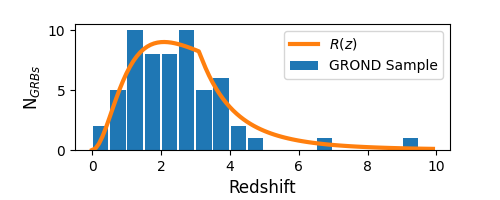}
                \caption{Overview of the sample of 88 published GRB afterglows observed by the GROND instrument with redshifts between $0.347 < z < 9.2$ used as baseline in this work. Top: The x-axis shows the time that GROND detected the GRB, and the y-axis represents the corrected $H$-band AB magnitude of the afterglow at the time of detection. The color of each data point represents the redshift of the event, where grey data points indicate GRBs for which a redshift determination was not made. Bottom: Histogram of the redshift of the 59 GRBs in the GROND sample with a measured redshift. The yellow curve plots the GRB rate function described by Equation \ref{eq:GRB Redshift Distribution} (with arbitrary normalisation to match the scale of the data).}
                \label{fig:GROND data overview}
                \end{center}
            \end{figure}

            To construct a simplified probability distribution for H-band magnitude we draw on a sample of 88 long GRB afterglows observed by GROND between 2007-2016 between redshifts $z \sim 0.3 - 9$. All GRBs in the sample were detected before rest frame $t = 4$ hours ($\sim 1.5 \times 10^4$ sec), and the data is recorded in the $K$-band and corrected into the $H$-band using $H = K + 0.20$ (which presumes the spectral energy distribution $F \propto \nu^{-\beta}$ with $\beta = 0.6$ as in the optical wavelengths \citep{Kann_2006, Greiner_2011}) to match the bandpass of our simulated near-infrared telescope (see Section \ref{subsubsec:SkyHopper}). An overview of the GRB detection times and corrected $H$-band magnitudes colour-coded by event redshift is shown in Figure \ref{fig:GROND data overview}, where the redshift distribution of the 59 events with measured redshift is consistent with the distribution described by Equation \ref{eq:GRB Redshift Distribution} (Figure \ref{fig:GROND data overview}, bottom panel).
            
            \begin{figure}[t!]
                \begin{center}
                \includegraphics[width=\columnwidth]{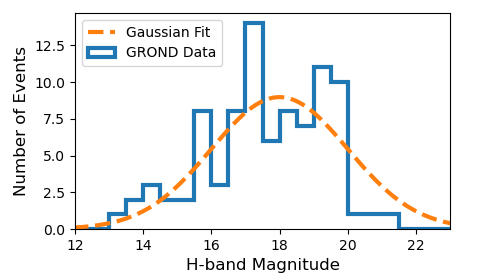}
                \caption{Rescaled $H$-band AB magnitude distribution from 88 GRB afterglows observed by GROND. Each GRB has been rescaled to a common time $t = 10^3$ sec post-burst (in the observer frame) using the light curves described in Section \ref{subsec: Afterglow Light Curve}.}
                \label{fig:Rescaled GROND Magnitudes}
                \end{center}
            \end{figure}
            
            In order to conveniently sample from this magnitude distribution, we rescale the magnitude of each event to the common observer frame time of $t_{\text{obs}} = 10^3$ seconds post-burst using the light curve described in Section \ref{subsec: Afterglow Light Curve} (Equation \ref{eq:light curve}) and fit the resultant magnitude distribution with a Gaussian distribution with mean $H_{AB} = 18$ and $\sigma = 2$ (Figure \ref{fig:Rescaled GROND Magnitudes}).
            
            To correct the magnitude of each GRB event to account for distance, k-correction, time dilation and host-galaxy extinction we undertake the following procedure: 
            
            We first generate a sample of $\sim 10^5$ afterglows each with a redshift (using Equation \ref{eq:GRB Redshift Distribution}) and H-band magnitude at $t = 1000$ sec (sampled as per Figure \ref{fig:Rescaled GROND Magnitudes}). For each event we calculate a correction to the observed magnitude resulting from distance ($D$), k correction ($k$) time dilation ($\tau$) and reddening effects, assuming that the magnitude sampled from the GROND distribution corresponds to a GRB at $z = 2$ (the mean GRB redshift\footnote{\url{https://www.mpe.mpg.de/~jcg/grbgen.html}}).
                % the mean magnitude corrections resulting from distance modulus ($\mu_{\text{D}}$), k-correction ($\mu_{\text{k-corr}}$) and time-dilation ($\mu_{\text{time dil.}}$) corrections. 
                
                \begin{align}
                    \Delta m_{D}(z) &= -5 \log \bigg( { \frac{ D_L(2) } { D_L(z) } } \bigg)  \\
                    \Delta m_{k}(z) &= (-\beta + 1) \cdot 2.5\log \bigg( { \frac{ 1 + 2 } { 1 + z } }\bigg) \\
                    \Delta m_{\tau}(z) &= \alpha \cdot 2.5\log \bigg( { \frac{ 1 + 2 } { 1 + z } }\bigg)
                \end{align}
                
                where we adopt the notation for observed GRB flux $F \propto \nu^{-\beta}t^{-\alpha}$ (with $\alpha = 0.85$ as in Section \ref{subsec: Afterglow Light Curve} and $\beta = 0.6$ as before), and $D_L(z)$ is the luminosity distance to the source.
                
                To simulate the effects of dust reddening as a function of redshift, we rescale the extinction curve of the Small Magellanic Cloud \citep{Gordon_2003} to arbitrary redshift $z$ by normalising it to the value of the UV host galaxy extinction at that redshift ($A_{M_{UV}}(z)$ from \citet{Trenti_2015}) at $\lambda = 0.14\mu$m. We then use the appropriately rescaled curves to subtract the extinction that would have occured if the burst had originated from redshift $2$, and then apply the extinction for a burst originating from redshift $z$.
                
                Using the sample of $10^5$ afterglows we calculate the mean magnitude correction $\mu$ due to distance, k-correction, time dilation and reddening effects. Then, in order to preserve the mean of the GROND distribution we apply a magnitude correction to each event proportional to the difference from the mean for each of these quantities:
                
                \begin{align}
                    \begin{split}
                        m_{\text{obs}} = m_{\text{sampled}} +  (\Delta m_{D}(z) - \mu_{D} &) \\
                        + (\Delta m_{k}(z) - \mu_{k}) + ( \Delta m_{\tau}(z) - \mu_{\tau} &) \\
                        + (\Delta m_{\text{reddening}}(z) - \mu_{\text{reddening}} &)
                    \end{split}
                \end{align}
                
                where we have calculated $\mu_{D} = 0.36$, $\mu_{k} = -0.06$ $\mu_{\tau} = -0.12$, $\mu_{\text{reddening}} = 0.03$ when sampling from the redshift distribution in Section \ref{subsec:Redshift Distribution}.

            % We adopt this method as it corrects the brightness of each GRB in accordance with its redshift (which is critical in estimating the number of high-redshift GRB detections) while also preserving the mean observed H-band magnitude in accordance with the GROND sample. 
            
            % This approach contains several problems: it ignores changes in the dust reddening across redshift, which reduces the brightness $z\gtrsim 6$ GRBs substantially less than $z\sim 2$ ones; it is limited by the completeness and inherent selection effects of the GROND afterglow sample; it neglects models of off-axis emmissivity; and it presumes consistent time evolution and spectral energy distribution behaviour for all GRBs and across time, among other issues. However, since a first-principles approach would be equally problematic due to the reasons outlined above, we judge that the above method is sufficient for the purpose of testing of the capabilities of an infrared nano-satellite in observing high-redshift GRB afterglows.

\section{Satellite Mission Scenario \& Telescope Parameters}
\label{sec:Satellite Mission Scenario & Telescope Parameters}
        
        This section outlines the relevant details taken into account when simulating the operations of the proposed near-infrared nano-satellite, as well as our approach to simulating the Swift space telescope and GROND instrument.
        
        \subsection{Nano-Satellite}
        \label{subsec:nano-satellite}
            
            Evaluating the performance of a nano-satellite in detecting GRB afterglows requires defining a specific mission scenario. This section will outline the SkyHopper\footnote{\url{https://skyhopper.research.unimelb.edu.au/}} nano-satellite concept which we will use as the baseline mission scenario for this study, focusing on the key mission parameters used in our analysis and the motivations behind them.
            
            % For this work we assume as baseline the SkyHopper mission concept (see Section \ref{subsubsec:SkyHopper}). The following section defines the key mission parameters used in our analysis and the motivations behind each of them.
            
            \subsubsection{SkyHopper Satellite Concept}
            \label{subsubsec:SkyHopper}
            
                The SkyHopper space telescope is a concept for a rapid-response near-infrared nano-satellite being developed by the University of Melbourne and several other Australian and international collaborators. 
                
                The spacecraft concept is based on a 12U CubeSat format ($\sim36 \times 22 \times 24$ cm$^3$), which will house a $20 \times 10$ cm$^2$ rectangular telescope mirror (0.15m equivalent aperture), feeding light to a $2048 \times 2048$ H2RG IR image sensor which is actively cooled to 145 K (a reference nominal operating temperature for near-infrared astronomical observations below $1.7\mu$m; e.g. see \citealt{Dressel_2021} for Wide Field Camera 3 on the Hubble Space Telescope). SkyHopper is designed to have a $\sim 1.5$ deg$^2$ FOV, and will be capable of maintaining an RMS pointing stability of $< 4"$ during observations.
                
                SkyHopper is designed to image in four filters across the near-infrared (0.8-1.7$\mu$m) simultaneously, with Swift UVOT-like celestial body avoidance angles (Table \ref{tab:Celestial Body Avoidance Angles}). The simultaneous four-band imaging capability allows the spacecraft to determine robustly a photometric redshift estimate for GRBs with $z \sim 6-13$ by measuring the Lyman break between two filters. This technique was first applied to the study of high redshift GRBs to estimate the redshift of GRB050904 at $z = 6.3$ \citep{Tagliaferri_2005} (in agreement with later spectroscopic measurements from the Subaru telescope; see \citealt{Kawai_2006}) and has been used on several occasions over the following years (e.g. see \citealt{Greiner_2009, Kruhler_2011, Cucchiara_2011}). Given the rapid decay in afterglow brightness it is critical that these observations are performed simultaneously to ensure that reduced flux in one filter can be unambiguously associated with absorption by neutral hydrogen rather than the natural decay in brightness of the source.
                
                \begin{table}[t!]
                    \caption{SkyHopper reference signal-to-noise ratios calculated for an m$_{AB}$ = 19.5 point source.}
                    \centering
                    \begin{tabular}{ c  c }
                        \hline\hline
                        Exposure Time & Reference \\
                        (mins) & SNR \\
                        \hline
                        1 & 1.1 \\
                        2 & 1.84 \\
                        3 & 2.42 \\
                        4 & 2.91 \\
                        5 & 3.34 \\
                        6 & 3.73 \\
                        7 & 4.08 \\
                        8 & 4.41 \\
                        9 & 4.71 \\
                        10 & 5 \\
                        \hline\hline
                    \end{tabular}
                    \label{tab:snr reference values}
                \end{table}
                
                The near-infrared detector on SkyHopper has a design goal to achieve point-source sensitivity of m$_{AB}$ = 19.5 [$t = 600s$; $5\sigma$; $H$-band], where the dominating noise contribution at $1.6\mu$m is assumed to be the thermal foreground from a relatively warm ($T \sim 250$ K) telescope baffle. For exposures longer than 10 minutes, the signal-to-noise ratio (SNR) of a single exposure can be approximated by the rescaling SNR $\propto \sqrt{t}$, however for shorter exposures this approximation breaks down as the readout noise becomes dominant\footnote{Poisson noise from source photons may dominate for very bright targets, but in this case the afterglow would be clearly detected, hence we can neglect to take it into account to determine marginal S/N detections.}. Within this domain we have computed a set of reference signal-to-noise values for a m$_{AB}$ = 19.5 point source with exposures ranging from 1 to 10 minutes, using an exposure time calculator for the SkyHopper telescope. Results are summarised in Table \ref{tab:snr reference values}.

            \subsubsection{Orbit Selection}
            \label{subsubsec:nano-satellite Orbit}  
            
                The small dimensions of a nano-satellite limit the amount of on-board space available for batteries, making it desirable for it to fly in sun-synchronous orbit as it allows the spacecraft to receive a constant stream of electricity via its solar panels and thus circumvents the need to store large amounts of power.
                
                For the analysis performed in this work, we model a nano-satellite in a Sun-synchronous dawn/dusk orbit:
                
                \begin{itemize}
                    \item altitude h = 550km;
                    \item circular orbit (eccentricity = 0);
                    \item dawn/dusk Sun-synchronous orbit (inclination = 97.6$^o$).
                \end{itemize}

            \subsubsection{Satellite TeleCommand and Slew Rate}
            \label{subsubsec:nano-satellite Telecommand and slew rate}
                
                Due to the power-law decay in afterglow brightness it is critical to understand how quickly a satellite can commence observations of the target, as the increased flux at early times gives the instrument a much higher probability of detecting the afterglow if it can arrive on target quickly. The two design aspects of the nano-satellite that are relevant to this calculation are the satellite's slew rate and the telecommunications latency in uplinking a re-pointing command to the spacecraft.
                
                The SkyHopper spacecraft is capable of slewing at a nominal rate of 2 deg s$^{-1}$. In order to calculate the time delay between re-pointing command and target acquisition we assign the satellite a random pointing (one which obeys bright-source avoidance angles) at the time of each burst and calculate the slew time as the angular difference between the current pointing and the target coordinates divided by the slew rate\footnote{In actual operations the nano-satellite would need to maintain bright-source avoidance angles throughout its slew, but within the simplified framework of this simulation a direct slew yields a comparable result.}.
                
                With regards to satellite TeleCommand there are several options to choose from when designing a space mission. Traditional Telemetry and TeleCommand (TTC) schemes use ground stations or government satellite relay networks such as NASA's Tracking and Data Relay Satellites (TDRS) to communicate in near real-time with satellites in orbit, but utilising this network is not only costly but requires the satellite to be equipped with a large antenna which often exceeds the volume/mass constraints on nano-satellite missions. Furthermore, a single ground station could only provide infrequent communications since it would only be seen $\sim$ twice per day by a satellite in low Earth Sun-synchronous orbit due to Earth's rotation.
                
                \begin{table}[b!]
                    \caption{Nominal Iridium TeleCommunications uplink latency for 550km sun-synchronous orbit (as per \citealt{Mearns_2018}). The `Probability' column indicates the probability that contact is established with the satellite by the given time.}
                    \centering
                    \begin{tabular}{c c}
                    \hline\hline
                        Time (mins) & Probability \\
                        \hline
                        Immediately & 31.8\% \\
                        1 & 41.4\% \\
                        2 & 50\% \\
                        6.6 & 68\% \\
                        10 & 74.2\% \\
                        25.4 & 95\% \\
                        40 & 99.7\% \\
                        \hline\hline
                    \end{tabular}
                    % \caption{Nominal uplink communications latency of the Iridium communications network sending a signal to a satellite in a 550km Sun-Synchronous orbit \citep{Mearns_2018}. To sample uniformly from this cumulative probability distribution we perform a linear interpolation between the data points presented in the table.}
                    \label{tab:Uplink Latency CDF}
                    \end{table}
                
                A viable solution for budget-limited nano-satellite missions is to leverage existing machine-to-machine orbital telecommunication networks to send messages to satellites in low Earth orbit (LEO). One drawback to this solution is that such networks were not designed for space applications, and thus there can be delays in uplinking a command to the spacecraft as it passes between gaps of the telecommunications satellite coverage beams\footnote{The sat-com beams generally overlap with no gaps on the Earth surface, but this is not the case in low-Earth orbit as constellations such as Iridium or Globalstar only orbit at $\sim 780-1400$km.}. For our afterglow detectability study, this acts effectively as a stochastic-like source of TeleCommand latency for the space mission, where the distribution of latency times depends on the communications network being used and the satellite's orbit.
                
                In this work we presume that the nano-satellite mission is budget limited and simulate TeleCommand using the Iridium machine-to-machine orbital telecommunications network, sampling directly from a statistical model of the nominal Iridium network performance derived by \cite{Mearns_2018} for a 550km Sun-Synchronous Orbit (Table \ref{tab:Uplink Latency CDF}).

        \subsection{Swift BAT}
        \label{subsec:Swift}
        
            \subsubsection{Prompt Gamma-Ray Detection}
            \label{subsubsec:Swift gamma ray detection}
            
                For simplicity we presume that Swift can detect every GRB that is not directly obstructed by the Earth, Sun or Moon. This deliberately ignores the fact that the prompt emission from the GRB must meet a certain flux threshold to be detected, and that BAT only has a 1.4sr FOV\footnote{ \url{ https://swift.gsfc.nasa.gov/about_swift/bat_desc.html} }. Ignoring these constraints artificially inflates the number of GRBs observable by Swift and hence the number of Swift triggers available for follow-up in a given year.
                
                This simplification however will not influence our overall results because they are presented as the ratio of GRBs detected divided by the number of Swift GRB triggers, which is agnostic to the total number of GRBs detected by Swift.
            
            \subsubsection{Orbit Details and Downlink Capabilities}
            \label{subsubsec:Swift Orbit and Downlink}
            
                We model the orbit of the Swift space telescope using its current orbital parameters at the time of writing:
                
                \begin{itemize}
                    \item altitude h = 561km;
                    \item circular orbit (eccentricity = 0);
                    \item inclination = 20.6$^o$.
                \end{itemize}
                
                The Swift Observatory utilises NASA's TDRS System to automatically send GRB triggers to the GRB Coordinates Network (GCN). With regards to its source localisation capabilities, upon detecting the prompt gamma-ray emission from a GRB, Swift's BAT is able to autonomously localise the event to within 1-3 arcmin, downlinking this information as a GRB trigger distributed via the GCN approximately 20 seconds after the initial detection \citep{Troja_2020}. Following this, Swift automatically re-points the telescope (providing the burst meets visibility constraints) to observe the afterglow with its XRT, refining the localisation to within a radius of a few arcsec \citep{Evans_2009} and distributing this via the GCN at approximately 100 seconds post-burst.
                
                For the purposes of this simulation, we assume that every GRB trigger takes exactly 20 seconds to downlink from Swift to Earth. The prompt BAT localisation of 1-3 arcmin is sufficient for the nano-satellite to reliably observe the GRB due to its 1.5 deg$^2$ FOV (as per SkyHopper's design requirements) which makes a shift in source position of 1-3 arcmin irrelevant. This means the nano-satellite can slew and commence observations of the source without the need to wait for Swift's enhanced GRB localisation.
                
        \subsection{GROND}
        \label{subsec:GROND}
            
            In order to compare space-based and ground-based near-infrared afterglow observations we employ a simplified model of the GROND instrument, taking into account the key points of difference between ground and space-based observations: sky visibility, weather and relative sensitivity at near-infrared wavelengths.
            
            \subsubsection{Modelling Ground-Based Observations}
            \label{subsubsec:Modelling Ground-Based Observations}
            
                GROND is one of the instruments on the MPG/ESO 2.2m telescope at La Silla observatory in Chile\footnote{\url{https://www.eso.org/sci/facilities/lasilla/telescopes/national/2p2/overview.html}}. Simulating GROND's operations requires taking into account not only the line-of-sight (LoS) to the target but also the time of day and the weather in Chile. 
                
                With regards to the local time of day we presume that GROND is only able to observe GRB afterglows during the hours of 8pm and 6am local time (UTC-4). While in reality these times shift seasonally throughout the year, we consider them to be a sufficient representation of `average' night time such that it will not substantially influence our results to neglect the seasonal variation.
                
                % To model the impact of weather on photometric observations we draw on weather data collected by the astronomers on duty at La Silla between Jan 1991 - May 1999\footnote{\url{https://www.eso.org/sci/facilities/lasilla/astclim/weather/tablemwr.html}} to estimate the probability that GROND will be able to make photometric observations on a given night (Figure \ref{fig:GROND weather modelling}). While this data varies substantially between months, using the mean value of P(clear night) $= 0.621$ is equivalent within the context our Monte Carlo simulation, which averages the results of thousands of GRB observations distributed uniformly throughout the year.
                
                To estimate the impact of weather on ground-based observations we draw on weather data collected by the astronomers on duty at La Silla between Jan 1991 - May 1999\footnote{\url{https://www.eso.org/sci/facilities/lasilla/astclim/weather/tablemwr.html}}, which records the fraction of nights each month where `photometric' observing conditions (no visible clouds, transparency variations under 2\%) were met, as well as the number of `useless' nights where the weather conditions meant that the telescopes had to be closed due to wind and/or humidity. Both of these metrics are inaccurate representations of the impact of the weather on GROND's operations - often GROND would make observations (at a reduced sensitivity) even when nights were not `clear' (less than 10\% of the sky covered in clouds, transparency variations under 10\%), let alone `photometric'. Conversely it is an overestimate to say that GROND was able to observe on every night where the telescope dome was open.
                
                Given the lack of more robust weather data and for the sake of comparing ground and space-based observing conditions we shall take these two scenarios - `photometric' observing conditions, and the telescope dome being open - as the lower and upper bounds on the impact of weather on GROND. Averaging the $\sim 8$ years of data recorded at La Silla, $62\%$ of nights are `photometric', and only $15\%$ are `useless' (i.e. $85\%$ are usable)\footnote{By definition, a night that is `photometric' will also be `usable'.}. For both scenarios we presume that GROND maintains the same point-source sensitivity described in Section \ref{subsubsec:GROND detector properties}, and that the classification of a given night will apply to the entirety of the night.
                
                % to estimate the probability that GROND will be able to make photometric observations on a given night (Figure \ref{fig:GROND weather modelling}). While this data varies substantially between months, using the mean value of P(clear night) $= 0.621$ is equivalent within the context our Monte Carlo simulation, which averages the results of thousands of GRB observations distributed uniformly throughout the year.
                
                % \begin{figure}[t!]
                %      \centering
                %      \includegraphics[width=\columnwidth]{figures/grond weather.png}
                %      \caption{Weather variability at La Silla observatory, Chile. The percentage along the y-axis reflects the probability that the weather will be clear enough on any given night to allow for photometric observations to take place, averaged over a one month timescale. The red line represents the mean probability over every month of data. (sourced from \url{https://www.eso.org/sci/facilities/lasilla/astclim/weather/tablemwr.html}).}
                %      \label{fig:GROND weather modelling}
                % \end{figure}
            
            \subsubsection{Telescope Properties and Detector Specifications}
            \label{subsubsec:GROND detector properties}
            
                GROND functions as part of the 2.2m MPG/ESO telescope, which it shares with both the Wide Field Imager and the Fiber-fed Extended Range Optical Spectrograph (FEROS) spectrograph. In the past, when a GRB trigger is picked up through the GCN, a mirror is folded into place (taking $\sim 20$ seconds) to re-direct the incoming light into the GROND instrument \citep{Greiner_2008}. GROND software would then autonomously re-point the telescope to the target as soon as it becomes accessible in the sky over Chile. Upon a night trigger, the software interrupts the running exposure while reading out the data, which depending on the instrument can take 10-45s. For simplicity, in this work we presume the entire instrument changeover and re-pointing of the observatory dome takes 120 seconds for every burst, meaning that the minimum possible time delay between Swift broadcasting a GRB trigger and GROND commencing its observations is 120 seconds (with extended time delays being related to the day/night cycle and LoS accessibility of the target).
                
                In our modelling of afterglow observations we presume that GROND has a $H$-band 5-$\sigma$ limiting magnitude of $H_{AB} = 20.1$ for an 8 minute Observation Block\footnote{\url{https://www.mpe.mpg.de/~jcg/GROND/operations.html}} (an 8 minute Observing Block typically consists of four $1.6$ minute exposures in the visual channels, and forty eight $10$ second exposures in the J, H and K bands). This value reflects the sensitivity of GROND observing under a new Moon with typical $1.0"$ of seeing and minimal airmass (1.0), and so it represents better than typical observing conditions from the ground.  
                Given the spatially and temporally fluctuating atmospheric foreground noise GROND's afterglow observations consist of a number of short exposures of the target which are stacked together. To simulate this behaviour we presume that GROND takes a number of 8 minute exposures of the target, and that the noise level is the same between exposures, and so $\text{SNR} \propto (\text{number of exposures})^{1/2} \propto (\text{exposure time})^{1/2}$. Therefore, if GROND observes a $H_{AB} = 20.1$ source for a total duration $t$:
                
                % Instead of simulating this behaviour directly we make the approximation that $\text{SNR} \propto (\text{exposure time})^{1/2}$, since infrared afterglows are faint sources and the dominant noise is from sky foreground (atmosphere).
                
                \begin{equation}\label{Eq:SNR_GROND}
                    \text{SNR}_{\text{ref}}(t)|_{\text{GROND}} = 5 \cdot \sqrt{ \frac{t}{8 \text{ mins}}}
                \end{equation}
                
                 %We then use Equation \ref{eq:magnitude to snr} to calculate the signal-to-noise ratio achieved by GROND for an exposure of arbitrary duration, taking $m_{\text{ref}} = 20.1$. 
                 In reality (i.e., taking into account the time-variable atmospheric foreground noise), Equation \ref{Eq:SNR_GROND} is somewhat over-estimating $\mathrm{SNR_{ref}}$ in the case of a long total exposure time, and under-estimating it for shorter exposures, but the overall impact on our analysis is not significant. 
                 
                In terms of scheduling observations we presume that GROND observes the afterglow for as long as possible on the first night after the GRB trigger, but that it does not schedule further observations for subsequent nights. Note that 
                %this behaviour reflects the operating mode of GROND after $2016$
                %JG: post-2016 GROND is only doing random shots
                until $2016$ every GRB was (weather permitting) observed with GROND also in the following nights.

\section{Afterglow Observation  Pipeline}
\label{sec:Observational Pipeline}
    
    This section presents an overview of our simulation approach, and details the methods we use for line-of-sight access calculations and to determine the signal-to-noise ratio of the afterglow photometry.
    
    \subsection{Simulation Strategy}
    \label{subsec:Simulation Strategy}
    
        To investigate the level to which a nano-satellite can perform follow-up observations on Swift GRB triggers we simulate the observational pipeline as follows:
        
        \begin{itemize}
        
            \item Using the models outlined in Section \ref{sec:Simulating Gamma-Ray Burst Events} we randomly generate GRB events assuming a uniform distribution on the celestial sphere, where each event occurs at a random time sampled from a uniform temporal distribution throughout the year. We also randomly generate the number of GRB events which occur each year, sampling this from a Poisson distribution centred on 380 GRBs yr$^{-1}$ (the average GRB detection rate of the Fermi Gamma-ray Burst Monitor if it could observe the whole sky \citep{von_Kienlin_2020}). For each GRB we randomly generate the $H$-band magnitude and the redshift of the event, where the redshift-corrected $H$-band magnitude of the afterglow at observer-frame $t = 10^3$ sec is generated as per Section \ref{subsec:Modelling Afterglow Emission}, and the redshift of each event is generated according to Equation \ref{eq:GRB Redshift Distribution}.
            
            \item For each GRB event, we determine whether Swift BAT is able to detect the GRB (Section \ref{subsec:Swift}). If it successfully detects the GRB, there is a 20 second delay in downlinking the GRB trigger to Earth using the NASA's TDRS network. Additionally, we determine whether the GRB is accessible for UVOT follow-up by checking whether Swift can achieve LoS access to the target within 90 minutes. If Swift BAT does not detect the GRB, no further action is taken.
            
            \item Upon receiving a GRB trigger in the nano-satellite control centre, a random uplink time and re-pointing time are sampled from their relevant probability distributions (Section \ref{subsubsec:nano-satellite Telecommand and slew rate}) and we calculate the earliest time that the satellite is able to achieve LoS access to the target. We perform the same computation for observations with the GROND instrument, excluding the uplink communications latency from the calculation, and taking into account the additional variables of the day/night cycle and the probability of the weather prohibiting observations.
            
            \item As soon as an observatory achieves LoS access to the target it begins making observations of the afterglow for the entire duration of the observing window, meaning that observations are only cut short either by LoS obstruction, or (for ground based observatories) dawn. The SNR of the entire observation is then calculated, taking into account the power-law decay of the afterglow brightness. Observations totalling SNR $> 5$ are considered a successful afterglow detection.
            
        \end{itemize}
    
    \subsection{Simulating Afterglow Observations}
    \label{subsubsec:Simulating Afterglow Observations}
    
        To calculate the flux captured in a single exposure we integrate the broken power-law light curve (Equation \ref{eq:light curve}) over the duration of the exposure (converting the magnitude of the GRB at the time of the observation into a reference flux in order to perform the integration), before converting the resultant fluence back into an observed AB magnitude `$m$'. The SNR for an exposure of duration $t$ is then computed from the reference SNR for the instrument considered, assuming that the signal scales as flux:
        %and that noise is independent of flux (i.e., neglecting source Poisson noise, which is justified due to the faintness of GRB afterglows):  
        %presuming that each exposure is read-out-noise dominated (which is the dominant source of noise for short exposures of faint sources in the infrared). In this case it follows that the signal-to-noise ratio for a given exposure can be calculated for a specific instrument using:
        
        \begin{equation}\label{eq:magnitude to snr}
            \text{SNR}(t) = \text{SNR}_{\text{ref}}(t) \cdot 10^{\textstyle (m_{\text{ref}}-m(t))/2.5}
        \end{equation}
        
        where $m_{\text{ref}}$ and SNR$_{\text{ref}}(t)$ are reference magnitudes and signal-to-noise values for an exposure of the same duration $t$. Note that (1) this equation assumes that the afterglow is faint, i.e., that the noise does not depend on the afterglow signal (e.g. background/foreground or detector noise limited); (2) these reference values are specific to the telescope making the afterglow observation; (3) the reference values and relevant noise sources for the near-infrared nano-satellite and the GROND instrument are discussed in Sections \ref{subsec:nano-satellite} and \ref{subsec:GROND} respectively.
        
        In the case where a telescope takes multiple exposures of the source each with their own individual durations $t_i$ the SNR from each exposure can be combined as:
        
        \begin{equation}\label{eq:snr combination}
            \text{SNR}\Big(n \text{ exposures}\Big) = \sqrt{ \quad \sum_{i=1}^{n} \text{ SNR}_i(t_i)^2 \quad}
        \end{equation}
        
        where this equation assumes that the noise is uncorrelated between exposures. In this work we consider a `detection' to be achieved when an afterglow observation reaches a cumulative SNR greater than 5.
        
    \subsection{Line of Sight Access Modelling}
    \label{subsec: Line of Sight Access Modelling}
        
        The opportunistic observation of stochastic GRB events relies on having LoS access to the target. We simulate the influence of this directly, designing our own LoS access code and implementing it as follows.
        
        We first generate a dataset of cartesian coordinates for each instrument relevant to our simulation - a nano-satellite, Swift and GROND - as well as the Sun, Earth and Moon (the most relevant astronomical bodies to Earth-based observations due to their apparent size and brightness). This dataset comprises of timestamped cartesian coordinates spanning the entirety of the year of 2023 (1 Jan 2023 00:00:00 UTCG - 1 Jan 2024 00:00:00 UTCG) in 20 second timesteps. For simplicity, when simulating multiple years of observations we simply repeat the year of 2023 multiple times.

        % \caption{Celestial body limb avoidance angles for each of the telescopes modelled in this work. The Sun avoidance angle is not relevant within our simulation for the GROND instrument as we simply presume that GROND can only observe during local night time (see Section \ref{subsubsec:Modelling Ground-Based Observations}). GROND's $20^{\degree}$ Earth-avoidance angle is equivalent to only observing 20$^o$ above the local horizon. Swift UVOT avoidance angles are taken from \url{https://swift.gsfc.nasa.gov/analysis/uvot_digest/numbers.html}}
        
        Using this set of cartesian coordinates we are able to determine the line-of-sight accessibility of an arbitrary point on the sky for each instrument, which enables us to determine the time that each instrument can access the GRB (after considerations of uplink and slew time are taken into account) as well as how long each instrument has to observe before the target coordinates become obstructed. To do so we take into account the radius and distance from the observer of each celestial body as well as the bright source avoidance angles for each telescope, which are quoted in Table \ref{tab:Celestial Body Avoidance Angles}. The operational constraints for Swift UVOT were taken from NASA's online UVOT digest\footnote{\url{https://swift.gsfc.nasa.gov/analysis/uvot_digest/numbers.html}}. A full discussion of each instrument listed in the table can be found in Sections \ref{subsec:nano-satellite} (Nano-Satellite), \ref{subsec:Swift} (Swift BAT/UVOT) and \ref{subsec:GROND} (GROND).

        \begin{table}[t!]
            \caption{Celestial body limb avoidance angles for each of the telescopes modelled in this work.}
            \centering
            \begin{tabular}{c  c  c  c }
                \hline\hline
                %  & \multicolumn{3}{ c }{\textit{Limb Avoidance Angle}} \\
                % \hline
                 & \multicolumn{3}{ c }{\textit{Avoidance Angle}} \\
                \cline{2-4}
                Satellite & Sun & Earth & Moon \\
                \hline
                Swift (BAT) & 0$^o$ & 0$^o$ & 0$^o$ \\
                Swift (UVOT) & 46$^o$ & 28$^o$ & 23$^o$ \\
                SkyHopper & 46$^o$ & 28$^o$ & 23$^o$ \\
                GROND & n/a & 20$^o$ & 9$^o$ \\
                \hline\hline
            \end{tabular}
            
            \label{tab:Celestial Body Avoidance Angles}
        \end{table}
        
\section{Optimising GRB Observations in Low Earth Orbit}
\label{sec:Optimising GRB Observations}

    When making early-time observations of GRB afterglows it is important to construct an exposure strategy that takes into account the power-law decay in the source brightness, since taking a single long exposure of the target risks washing out the increased signal from the bright early-time afterglow with background noise. It is possible to achieve a higher signal-to-noise observation by taking short exposures of the source at early times to capitalise on the increased source brightness before taking longer exposures as the power-law decay begins to flatten off. We use Monte Carlo methods to compare the performance of several different exposure strategies in order to identify the optimal strategy for use on board a rapid-response space telescope.
    
    \subsection{Optimising an Afterglow Exposure Strategy}
    \label{subsec:Optimising an Afterglow Exposure Strategy}
    
        We define an exposure strategy to be a sequence of consecutive exposures performed by the near-infrared nano-satellite, where for simplicity each exposure lasts for an integer number of minutes. 
        
        Presuming that the nano-satellite can achieve LoS access to the source, the amount of time available to observe the GRB afterglow depends on the satellite's position in orbit: it can be as short as a few seconds (if the GRB is about to be eclipsed by an astronomical body) and as long as $\sim 45$ minutes (this depends on the bright source avoidance angles - for SkyHopper in LEO this corresponds to slightly less than half an orbit). As such, the exposure strategies we explore must be able to vary their total duration to match the duration of the observing window. Throughout this work we express exposure strategies as ordered lists where each number represents the duration of a given exposure in minutes. For example: 
        
        \begin{itemize}
            \item A strategy that performs a 1 minute exposure followed by 2 and 3 minute exposures is notated as `$[1, 2, 3]$'.
            \item To construct variable length observing strategies we simply repeat the final exposure in the sequence for as long as the observing window allows. In our notation this is indicated with an ellipse, where the ellipse follows the exposure which will be repeated. For example, a strategy which consists of a 1 minute exposure followed by consecutive 5 minute exposures is notated as `[1, 5, 5, ...]'.
        \end{itemize} 
        
        The performance of each exposure strategy is judged with respect to two key criteria: the overall detection fraction and the speed at which it can make a detection. 
        
        Specifically, we focus on the detection of those GRBs which are also observable by Swift's UVOT\footnote{Here we wish to make a distinction between `observable' (meaning the telescope is able to safely point to, and take exposures of, the target coordinates) and `detectable' (meaning the observed flux from the afterglow is bright enough to be detected by the instrument). This distinction is critical, because while high redshift targets can be observable by Swift, they are not detectable due to the Lyman-$\alpha$ dropout, which is precisely what enables a photometric redshift estimation.}, so that it would be possible to distinguish between high redshift and `Dark' GRBs by comparing the near-infrared photometry from SkyHopper with UV/optical data from Swift. While we do not directly model the Lyman-$\alpha$ dropout for high redshift bursts, we assume that all bursts with $z > 5$ will not be detected by Swift, and therefore a detection in the near-infrared would confirm the GRB as originating from high redshift.
        
        For ease of reference we refer to a `UVOT-observable' Swift trigger as one which can be observed by Swift's UVOT within 90 minutes (approximately one Swift orbit) of the GRB prompt emission (i.e., outside the Earth, Sun and Moon limb avoidance angles cited for Swift (UVOT) in Table \ref{tab:Celestial Body Avoidance Angles}). This does not mean that the burst was necessarily detected by UVOT - simply that UVOT could safely re-point to the burst without violating celestial body avoidance angles.
        
        The detection fraction of a given strategy is defined as:
        
        \begin{equation}\label{eq:Follow Up Detections}
            \text{Det. Fraction} = \frac{\text{ Follow-Up Detections }}{\text{UVOT-observable Triggers}}
        \end{equation}
        
        For ease of comparison GROND is also evaluated on the same definition of detection fraction.
        
        \subsubsection{Exposure Strategy Constraints}
        \label{subsubsec:Exposure Strategy Constraints}
        
            If one had access to limitless computing power and uplink TeleCommand capability then an afterglow exposure strategy could be tailored dynamically to observe each GRB, taking into account the time delay between the GRB occurring and observations commencing as well as the duration of the observing window. However, given the limitations in on-board computing power and the benefits in software development and testing from limiting TeleCommand complexity, we adopt a simplified approach in constructing an observing strategy: we presume that the exposure strategy is hard-coded into the satellite and the ground can only command the satellite to take an integer number of exposures from that strategy. One consequence of this approach is that the satellite will rarely be able to observe for the full duration of the exposure window (e.g. if an observing strategy consists of repeated 10 minute exposures and the observing window is 19 minutes long, the satellite would only be able only perform a single 10 minute exposure of the target).
            
            A key consideration when designing an near-infrared observing strategy for use in low Earth orbit are cosmic rays (CRs) and `snowball' events, which can compromise an exposure by depositing a large amount of charge onto the detector after a collision. Instead of directly simulating the impact of these events we simply place an upper limit of 15 minutes on any single exposure duration: CR events occur at a rate of $11 \pm 5$ CR/s for Hubble's WFC3/IR and impact 1-10 pixels on the detector \citep{Dressel_2021} meaning that any given pixel has a $< 5\%$ probability of being impacted by a cosmic ray for a 15 minute exposure assuming the same sized near-infrared detector\footnote{SkyHopper's design is based on a similar detector}. Considering that MgCdTe detectors such as Hubble's WFC3 and SkyHopper's can be read non-destructively multiple times during a single exposure, offering further mitigation against cosmic rays via fitting of the charge accumulation in a pixel versus time, we judge this impact as sufficiently low to not affect significantly our nano-satellite's ability to detect GRB afterglows, which only occupy a few pixels on the detector given they are point sources. The same argument applies to `snowball' events, which occur at a much lower rate of 1/hour \citep{Green_2020}. We also note that with these assumptions the impact of cosmic ray hits is smaller than or comparable to the probability that a GRB afterglow is along the same line of sight as a brighter foreground source (e.g. Galactic star), which is again estimated to be at the few percent level based on 2MASS number counts (estimating that a typical point source will occupy a few tens of pixels on the detector, and taking into account the density of point sources brighter than $H_{AB} = 20$ is of the order $\sim 10^3$ deg$^{-2}$ \citep{Skrutskie_2006}), and affects equally all observatories irrespective of their location on the ground or in space (note that brighter point sources will have diffraction spikes covering more pixels, but their number density in the sky drops more rapidly than the increase in the area covered, so for the estimate we can consider point sources of brightness comparable to the faintest afterglow we aim to detect). 
            
            \subsubsection{Identifying an Optimal Observing Strategy}
            \label{subsubsec:Identifying an Optimal Observing Strategy}
                
                We use a Monte Carlo approach to identify the optimal exposure strategy, testing each strategy on $\sim 10^5$ trial afterglow observations across 1000 years of simulated observations. For each trial we record the overall GRB detection fraction as well as the time that the GRB was detected.
                
                All of the results presented in this section (and those in Section \ref{sec:Results and Discussion}) were generated for two cases of early-time $t < 500$ sec light curve decay index (as per Section \ref{subsec: Afterglow Light Curve}): a flat light curve, and a rising light curve $F \propto t^{-\alpha}$, $\alpha = -0.2$. We found that the choice of early-time light curve had negligible impact on the detection statistics of a given strategy, and so here we present only the results from our `flat light curve' simulations.
                
                \begin{figure}[b!]
                    \begin{center}
                        \includegraphics[width=\columnwidth]{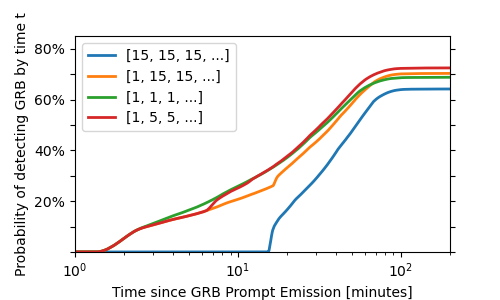}
                        \caption{Cumulative probability of making a follow-up detection on a UVOT-observable Swift GRB trigger for four different observing strategies, where $t = 0$ is the time the burst is detected by Swift BAT. Data is generated by simulating $\sim 10^5$ trial afterglow observations with each strategy.}
                        \label{fig:Detection CDF}
                    \end{center}
                \end{figure}

                Figure \ref{fig:Detection CDF} plots the cumulative probability that the nano-satellite will detect a Swift GRB trigger as a function of time after the trigger. This represents not only how quickly a given strategy can achieve a 5$\sigma$ observation but also takes into account all stochastic sources of time delay in the observing pipeline: Swift's downlink time, the TeleCommand latency when uploading a re-pointing command to the satellite, the time associated with waiting for LoS access to the target and the satellite's slew time.
                
                The worst performing strategy both in terms of timeliness and overall detection probability is one which takes consecutive 15 minute exposures of the target. While one might expect longer exposures to yield more signal and thus a higher probability of detecting the afterglow, given LoS visibility constraints in low Earth orbit a strategy which performs shorter exposures can observe the target for longer on average (i.e., if the observing window is 14 minutes long, a [5, 5, 5, ...] strategy can observe the target for a total of 10 minutes while a [15, 15, 15, ...] strategy cannot observe the target at all).
                
                The strategy which maximises the probability of an early-time ($t < 10$ minutes) GRB detection is one which makes consecutive 1 minute exposures of the target since it is able to make the best use of the increased early-time source flux. This strategy is also able to achieve a very high ($\sim 70\%$) overall probability of detecting the afterglow. However in terms of overall detection capability it is marginally less effective than strategies that include longer exposures, which are better suited to late-time detections.
                
                The best performing strategies at both early and late times are those which start with a 1 minute exposure of the target then transition to increasingly longer exposures (a strategy which is similar to the approach taken by GROND). Comparing specifically the [1, 5, 5, ...] and [1, 15, 15, ...] strategies, we find that the increased signal-to-noise afforded by longer exposures mean that the [1, 15, 15, ...] strategy is able to detect almost as many GRBs overall as the [1, 5, 5, ...] strategy ($\sim 2\%$ fewer). However, by utilising shorter exposures the [1, 5, 5, ...] strategy is able to make detections much more quickly, with its cumulative detection probability being consistently $\sim 2-5\%$ higher than the [1, 15, 15, ...] strategy for $t > 8$ minutes. 
                
                It is worth noting that within our simulation, performing additional 1-minute exposures at the start of an observing strategy would increase the probability of an intermediate-time detection ($t = 3-7$ minutes) by $\sim 5\%$ without substantially decreasing the overall detection probability. However, in this work we avoid over-tuning an observing strategy to maximise its effectiveness during this time period due to our simplifications in modelling the early-time GRB light curve (Section \ref{subsec: Afterglow Light Curve}). 
                
\begin{figure}[t!]
                    \begin{center}
                        \includegraphics[width=\columnwidth]{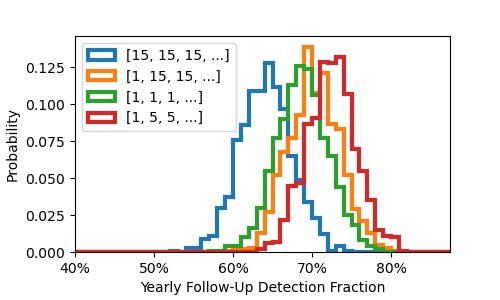}
                        \caption{PDF of the yearly follow-up detection fraction for four different observing strategies. The x-axis represents the fraction of UVOT-observable Swift triggers which were detected using the given exposure strategy. Data is generated by binning the results of $\sim 10^5$ trial afterglow observations into each year of simulated observations.}
                        \label{fig:Detection Fraction PDF}
                    \end{center}
                \end{figure}

                Figure \ref{fig:Detection Fraction PDF} shows the variability in yearly follow-up detection fraction for several different observing strategies. We find that this variability is the same regardless of strategy choice (standard deviation $\sigma \sim 3\%$), indicating that the spread comes from the intrinsic variability of the GRB afterglows and/or stochastic visibility constraints in low Earth orbit. Therefore, we can identify the optimal observing strategy purely on the grounds of detection timeliness and mean overall detection fraction.
                
                \begin{table}[b!]
                \caption{Total percentage of successful follow-up detections for each strategy investigated in this work divided by the number of UVOT-observable Swift triggers (see Section \ref{subsec:Optimising an Afterglow Exposure Strategy} for a description of the strategy notation).}
                    \centering
                    \begin{tabular}{c  c c }
                        \hline\hline
                         & \multicolumn{2}{c}{Detection Probability (\%)} \\
                        Strategy & All GRBs & z > 5 \\
                        \hline
                        \text{[}1, 1, 1, ...\text{]}    & $68.8 \pm 3.3$ & $39.3 \pm 11.9$ \\
                        \text{[}1, 5, 5, ...\text{]}    & $72.5 \pm 3.1$ & $44.1 \pm 12.3$ \\
                        \text{[}1, 15, 15, ...\text{]}  & $70.3 \pm 3.2$ & $41.7 \pm 12.2$ \\
                        \text{[}15, 15, 15, ...\text{]} & $64.2 \pm 3.3$ & $39.9 \pm 12.1$\\
                        \hline\hline
                    \end{tabular}
                    \label{tab:Strategy Analytics}
                \end{table}
                
                Table \ref{tab:Strategy Analytics} shows the total fraction of UVOT-observable Swift GRB triggers successfully detected every year by a  near-infrared nano-satellite employing a given strategy. In addition to those strategies listed in the table we made efforts to search for higher performing, more complex strategies such as a $[1, 3, 5, 5, ...]$ strategy, but found that their detection statistics differed negligibly from the $[1, 5, 5, ...]$ strategy. Therefore, we identify [1, 5, 5, ...] as the preferred GRB observing strategy; not only does it have a high probability of an early-time afterglow detection, but it also achieves the highest probability of detecting the afterglow overall. We find that taking 5 minute exposures of the GRB afterglow is more effective than taking 15 minute exposures, likely because on average the telescope is able to spend more time taking exposures of the afterglow given the variability in the length of the observing window from low Earth orbit. Additionally, from a mission design standpoint it is desirable to perform several shorter exposures in place of one long exposure to minimise the risk of a cosmic ray event disrupting an observation. In this work we elect not to further optimise the duration of the repeating exposure, leaving this to a future work which incorporates a more sophisticated model for GRB light curves.
        
\section{Results and Discussion}
\label{sec:Results and Discussion}

    This section outlines the numerical results generated by performing $\sim 10^5$ trial observations of GRB afterglows within our Monte Carlo simulation framework. We present the results of a near-infrared nano-satellite using the [1, 5, 5, ...] exposure strategy for observing GRB afterglows (as per Section \ref{sec:Optimising GRB Observations}), and for the GROND instrument using the observing strategy defined in Section \ref{subsec:GROND}.
    
    \subsection{Comparing Ground and Space-Based Observations}
    \label{subsec:Comparing Ground and Space-Based Observations}

\begin{figure}[t!]
            \begin{center}
                \includegraphics[width=\columnwidth]{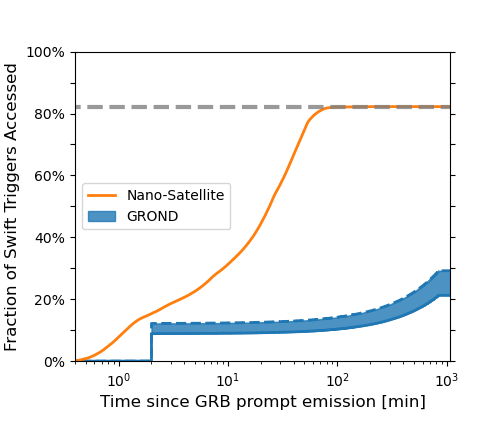}
                \caption{Comparison of the cumulative probability that the instrument begins taking exposures of an arbitrary Swift GRB trigger within a certain time for the GROND instrument and a nano-satellite in a 550km sun-synchronous orbit. The solid and dashed blue lines represent the worst case (`photometric') and best case (`usable') weather conditions on the ground, and the shaded blue region represents the possible range of GROND's performance. The jump in GROND's access fraction at $t = 120$ seconds is due to our simplified modelling assumption that GROND takes exactly 2 minutes to re-position its dome and begin observations on a target (see Section \ref{subsubsec:GROND detector properties}). The grey dashed line indicates the fraction of Swift GRB triggers that are UVOT-observable.}
                \label{fig:GROND access time comparison}
            \end{center}
        \end{figure}

        \begin{table}[b!]
            \caption{Number of events accessed/detected (in the $H$-band) divided by the number of UVOT-observable Swift GRB triggers for each year (\%). The third row represents the percentage of $z > 5$ events detected divided by the total number of UVOT-observable $z > 5$ events which occurred each year. The labels `phot.' (photometric) and `usable' denote the weather modelling assumptions used for GROND, which are described in Section \ref{subsubsec:Modelling Ground-Based Observations}.}
            \centering
            
            % \begin{tabular}{c  c  c }
            %     \hline\hline
            %     Instrument & \% Accessed & \% Detected \\
            %     \hline
            %     Nano-Satellite & $100.2 \pm 0.1$ & $72.9 \pm 3.1$\\
            %     GROND$_H$ (phot) & $26.0 \pm 3.1$ & $20.6 \pm 2.7$ \\
            %     GROND$_H$ (usable) & $26.0 \pm 3.1$ & $20.6 \pm 2.7$ \\
            %     \hline\hline
            % \end{tabular}
            
            \begin{tabular}{c c c c}
                \hline\hline
                 & Nano-     & GROND$_H$ & GROND$_H$\\
                 & Satellite & (phot.)    & (usable) \\
                 \hline
                 Accessed           & $100.3 \pm 0.1$ & $26.1 \pm 3.3$ & $35.9 \pm 3.8$  \\
                 Detected           & $72.5 \pm 3.1$  & $19.9 \pm 2.9$ & $27.3 \pm 3.3$  \\
                 $(z > 5)$ Det.     & $44.1 \pm 12.3$ & $13.4 \pm 8.9$ & $18.1 \pm 10.0$ \\
                 \hline\hline
            \end{tabular}
            
            % \caption{Total number of events accessed (or detected) by each instrument in the $H$-band divided by the number of UVOT-observable Swift GRB triggers for each year, collected from 1000 simulated observing runs (with 1$\sigma$ uncertainties). For GROND, this figure represents the number of GRBs it is able to detect/access within the first night after the GRB trigger (i.e., within 14 hours), whereas data from the infrared nano-satellite is taken from the first 2 hours after the burst. The nano-satellite is capable of accessing more than 100\% of UVOT-observable Swift triggers since its polar orbit enables access to a larger fraction of the sky than Swift's equatorial orbit.}
            \label{tab:Instrument Comparison}
        \end{table}
        Figure \ref{fig:GROND access time comparison} demonstrates the clear advantage that space-based afterglow observatories have over their ground-based counterparts: the nano-satellite can access every UVOT-observable Swift GRB within $\sim 1$ hour of the burst whereas GROND is only able to access $\sim 30\%$ of such bursts after 14 hours (under the assumption of optimal weather conditions).
        
        Within our simulation, the nano-satellite is capable of accessing more than 100\% of UVOT-observable Swift triggers (Table \ref{tab:Instrument Comparison}), meaning that there are GRBs which are inaccessible to UVOT but can be accessed by the nano-satellite. While the nano-satellite and UVOT share the same celestial body avoidance angles, UVOT's sky visibility is limited by its equatorial orbit (which passes between Earth and the Sun for a fraction of its orbit, obscuring two large sections of the sky simultaneously) whereas the sun-synchronous polar orbit of the nano-satellite affords it access to a much larger fraction of the sky throughout its orbit. Conversely a ground-based facility is limited by the weather, the day/night cycle and the limited fraction of sky accessible overhead, resulting in GROND only being able to access a relatively small fraction of Swift GRB triggers.

        % \begin{figure}[b!]
        %     \centering
        %     \includegraphics[width = \columnwidth]{figures/One SkyHopper Event Overview.png}
        %     \caption{Overview of every GRB observed by the near-infrared nano-satellite over 2 years of simulated observations. The x-axis shows the time that the near-infrared nano-satellite accessed the GRB, and the y-axis represents the $H$-band AB magnitude of the afterglow at the time of access. Circles indicate detections (using the $[1, 5, 5, ...]$ strategy), and crosses indicate non-detections.}
        %     \label{fig:SkyHopper Detection Overview}
        % \end{figure}
        \begin{figure*}[hbt!]
            \centering
            \includegraphics[width=0.45\textwidth]{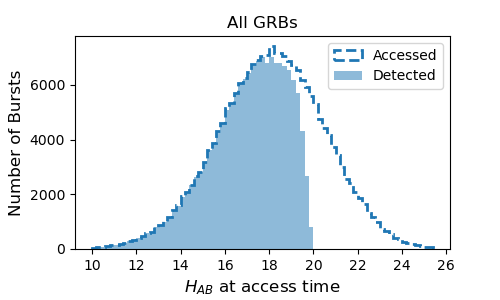}
            \includegraphics[width=0.45\textwidth]{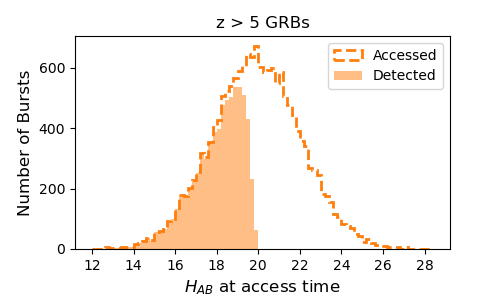}
            \caption{Histograms demonstrating the H-band afterglow magnitude of GRB afterglows at the time it was accessed by the near-infrared nano-satellite. The left plot depicts the full sample of $\sim 10^5$ afterglows, while the right plot shows only those events with $z > 5$. The dashed line indicates all afterglows generated in the simulation, and the shading indicates the GRBs which were detected by the near-infrared nano-satellite using the $[1, 5, 5, ...]$ strategy.}
            \label{fig:skyhopper burst histograms}
        \end{figure*}
        Despite having a slightly lower $H$-band point-source sensitivity, a nano-satellite with the SkyHopper design requirements is comparable to GROND at detecting GRBs in the $H$-band given its ability to acquire the targets more rapidly (the nano-satellite detects $\sim 72\%$ of the events that it accesses while GROND detects $\sim 76\%$ - Table \ref{tab:Instrument Comparison}). It is important to note however that this comparison only takes into account $H$-band detections - GROND's photometry from the bluer $g'$, $r'$, $i'$, $z'$ and $J$ band are more sensitive and are able to increase the likelihood of detecting a GRB afterglow at redshift $z\lesssim 8$, but this modelling aspect is outside the scope of this work. For this reason, these numbers should not be taken to reflect the total number of GRBs that each instrument will observe per year. However, given $H$-band detections are critical for the identification of GRBs at the highest redshift, this comparison demonstrates the advantage that a near-infrared nano-satellite has in identifying GRBs during the epoch of reionisation compared to a ground based observatory which is limited by the atmospheric thermal foreground. The advantage of space-based observatories would be even greater for observations in the $K$-band and at longer wavelengths, although in that case it would be challenging to meet thermal management requirements within a CubeSat form factor. 
        
        Figure \ref{fig:GROND access time comparison} highlights the fact that UVOT-observable triggers only account for $\sim$ 80\% of the GRBs detected by Swift in our simulation. In reality, UVOT is able to follow up a higher fraction of BAT triggers since Swift generally tries to avoid pointing BAT close to the exclusion zone around the Sun in order to maximise the number of GRBs that it can follow up with its XRT and UVOT. Our simulation under-predicts this fraction since we model BAT as having a full-sky field of view and being capable of detecting bursts arbitrarily close to the Sun. 
        
        Figure \ref{fig:skyhopper burst histograms} compares the intrinsic distribution of H-band magnitudes at the time of access by the near-infrared nano-satellite with those events that were detected across the $10^5$ trials. We find that the near-infrared nano-satellite is able to reliably detect bursts that have a H-band afterglow magnitude $H_{AB} \lesssim 19$ after re-pointing. This is true both for the full sample of afterglows as well as for the high redshift population, indicating that the primary reason for the reduced effectiveness in detecting high redshift bursts is because this population is on average $\sim 2$ magnitudes fainter. We find that the near-infrared nano-satellite is able to detect every GRB with $H_{AB} < 17$ and that detection efficiency drop off gradually between $17 < H_{AB} < 20$, which is likely due to only having a short window of time to observe the burst before it is obstructed by the Earth, Sun or Moon. 
        
    \subsection{High Redshift GRB Identification}
    \label{subsec:High Redshift GRB Detection Rate}
        
        From our simulation we find that a near-infrared nano-satellite using the $[1, 5, 5, ...]$ observing strategy is able to detect $44.1\% \pm 12.3\%$ of the UVOT-observable GRBs which originate at $z > 5$ (Table \ref{tab:Instrument Comparison}). To calculate the number of real-world detections this corresponds to, we calibrate the results from our simulation to the actual number of GRBs observed by Swift each year (our simulation over-predicts this number due to simplified modelling of Swift's observing pipeline - see Section \ref{subsubsec:Swift gamma ray detection}). Presuming that Swift observes an average of 76.2 GRBs yr$^{-1}$ with UVOT\footnote{Data taken from the Swift catalog between 2005-2019 selecting only those bursts with UVOT observations.}, and that number of GRBs observed by Swift which have redshift $z > 5$ is approximately $f_{z>5} \sim 5\%$ \citep{Wanderman_2010, Greiner_2011, Perley_2016} we can use the results from our simulation to compute the expected number of high-z GRBs observed by our nano-satellite as:
        
        \begin{align*}\label{eq:High z Detections}
            N_{\text{GRBs}}(z > 5) &= 76.2 \times f_{z>5} \times (44.1\% \pm 12.3\%)
            \\
            &= 1.68 \pm 0.47
        \end{align*}

         Therefore, we find that our near-infrared nano-satellite is able to detect $1.68 \pm 0.47$ high redshift GRBs per year to the $1\sigma$ confidence level. The uncertainty on this number in any given year is of course slightly higher, as there is Poisson uncertainty on the number of UVOT-observable Swift triggers that occur each year, which is not included in the above calculation.
        
        For completeness we apply this same method to the nano-satellite's overall detection fraction, and calculate that a near-infrared nano-satellite should be able to detect on average $\sim 55 \pm 2$ GRB afterglows yr$^{-1}$ (excluding again Poisson noise on Swift triggers). 
        
        It is worth noting that these figures represent conservative estimates of the number of GRB detections per year by a near-infrared nano-satellite, as they presume that the telescope only performs follow-up on Swift GRB triggers. In reality, there are several other current and future all-sky observatories that could increase the yearly number of GRB triggers available for follow-up. Such instruments include the existing Fermi Gamma-Ray Space Telescope which detects $\sim 240$ GRBs per year \citep{von_Kienlin_2020}, as well as several future instruments including the Franco-Chinese SVOM Mission which is expected to detect $\sim 60-70$ bursts per year \citep{Wei_2016}, or nano-satellite missions such as the CAMELOT \citep{Werner_2018} and HERMES \citep{Fiore_2020} projects. The only limitation to performing follow-up observations on GRB triggers from different instruments is that the source must be localised precisely enough to be reliably observed using a telescope with a relatively small field of view (SkyHopper's design requirements have a 1.5 deg$^2$ FoV).

    \subsection{Application of a Nano-Satellite Constellation}
    \label{subsec:Application of a Nano-Satellite Constellation}

        As an extension to the results presented above, we test the afterglow follow-up capabilities of a small fleet of identical near-infrared nano-satellites. We presume that each satellite orbits in the same orbital plane as defined in Section \ref{subsubsec:nano-satellite Orbit}, and are spaced uniformly around the orbit. 
        
        % \begin{figure}[b!]
        %     \begin{center}
        %         \includegraphics[width=\columnwidth]{figures/constellation nominal uplink.png}
        %         \caption{Cumulative probability of detecting a GRB afterglow when using a constellation of rapid-response infrared nano-satellites communicating using the Iridium telecommunications network. The detection time represents the earliest time that a $5\sigma$ observation is achieved by any individual satellite in the constellation. Data is recorded from 1000 iterations of a typical year of operations.}
        %         \label{fig:constellation cdf}
        %     \end{center}
        % \end{figure}
        
        To simulate observations, a random initial pointing and uplink communications delay is generated independently for each satellite in the constellation, and each satellite begins observations as soon as it can safely slew to the afterglow coordinates. Each satellite uses the [1, 5, 5, ...] strategy to observe GRBs, and the earliest detection time is recorded between all elements in the constellation. 
        
\begin{figure}[t!]
            \centering
            \begin{tabular}{@{}c@{}}
                \includegraphics[width=\columnwidth]{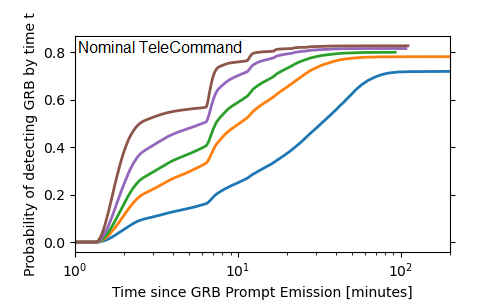}
            \end{tabular}
                
            \vspace{\floatsep}
            
            \begin{tabular}{@{}c@{}}
                \includegraphics[width=\columnwidth]{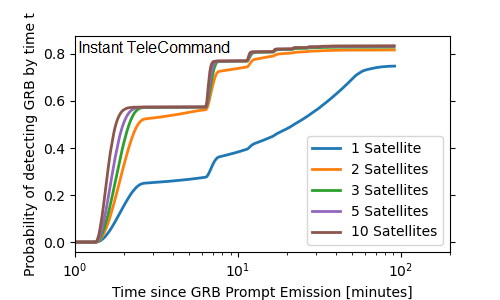}
            \end{tabular}
            
            \caption{Cumulative probability of detecting a GRB afterglow when using a constellation of rapid-response near-infrared nano-satellites. Top: Satellite TeleCommand is modelled using the nominal telecommand latency of the Iridium network (Section \ref{subsubsec:nano-satellite Telecommand and slew rate}). Bottom: Satellite TeleCommand is modelled as occuring instantaneously (no delay in uplinking a re-pointing command to any satellite in the constellation). The detection time represents the earliest time that a $5\sigma$ observation is achieved by any individual satellite in the constellation.}
            \label{fig:constellation cdf}
        \end{figure}

        With multiple satellites observing the same afterglow it becomes possible to stack the exposures from each satellite together to achieve an observation with higher signal-to-noise ratio overall, increasing the probability of detecting the source. We presume that the process of downlinking each of the images taken by each satellite in the constellation and stacking them would take several hours (each satellite must finish its full course of observations and downlink the final image over a limited telecommunication network) and so we do not include its effects in the prompt detection capabilities of the constellation (i.e., we exclude this effect from Figure \ref{fig:constellation cdf}), instead presenting the results of image stacking separately in Table \ref{tab:Total Multihopper Stacking}. We approximate the signal-to-noise increase of exposure stacking using Equation \ref{eq:snr combination} - in reality this process would be more complex and require inter-satellite calibration measures, but such modelling is beyond the scope of this work.
        
        % \begin{table}
        %     \caption{Total fraction of successful follow-up detections for different sizes of nano-satellite constellations both in the case of nominal (Columns 1-2) and instantaneous (Columns 3-4) TeleCommand. The results both before and after image stacking are shown for each case.}
        %     \centering
        %     \begin{tabular}{| c || c | c || c | c |}
        %         \hline
        %         \parbox[t]{2mm}{\multirow{4}{*}{\rotatebox[origin=c]{90}{ \textbf{\# Sats } }}} & \multicolumn{4}{|c|}{\textbf{Afterglow Detection Probability}}\\
        %         \cline{2-5}
        %          & \multicolumn{2}{|c||}{\textit{Nominal TeleCmd.}} & \multicolumn{2}{|c|}{\textit{Instant TeleCmd.}} \\
        %         \cline{2-5}
        %          & Before   & After    & Before   & After \\
        %          & Stacking & Stacking & Stacking & Stacking \\
        %         \hline\hline
        %         1  & 72.5\% & n/a & 74.82\% & n/a \\
        %         2  & 78.1\% & 79.7\% & 81.7\% & 82.9\% \\
        %         3  & 80.0\% & 82.6\% & 82.9\% & 85.4\% \\
        %         5  & 81.5\% & 86.0\% & 83.3\% & 88.1\% \\
        %         10 & 82.7\% & 89.6\% & 83.4\% & 91.0\% \\
        %         \hline
        %     \end{tabular}

        %     \label{tab:Constellation Stacking}
        % \end{table}

        Figure \ref{fig:constellation cdf} (upper panel) compares the cumulative probability of detecting a GRB over time for a range of constellation sizes. We find that in the case of nominal TeleCommand delay, launching a second satellite positioned on the other side of the globe doubles the probability of detecting a GRB at early times (t $\sim 3-11$ minutes), which is unsurprising given the largest constraint on timely GRB observations in LEO is prompt LoS visibility to the source. However, launching a second satellite does not drastically increase the overall probability of detecting a GRB target, yielding a $\sim 6\%$ increase in total detection fraction even after combining the signal from both two satellites. This behaviour is also true for larger constellations - making the constellation larger has a substantial impact on the early-time GRB detection capability without dramatically increasing the overall probability of detection. This indicates that the primary limitation on the total detection fraction of the constellation not its sky coverage, but rather the point-source sensitivity of their instruments. Even an increase of the collecting area for the orbiting satellite would still miss the faint-end tail of the afterglow luminosity distribution.
        
        Increasing the size of the nano-satellite constellation yields diminishing returns with regards to the early-time detection probability, since it only takes three satellites equally spaced around the globe to achieve 100\% sky coverage. To illustrate this point, we simulated the behaviour of a satellite constellation that was able to communicate without any uplink latency (Figure \ref{fig:constellation cdf}, lower panel). We find that if one can achieve zero-latency uplink to each satellite, a 3-satellite constellation is able to effectively match the performance of a 10 satellite constellation with only a short $\sim 60$ second delay in timeliness due to longer average re-pointing times for the 3 satellite constellation. 
        
        \begin{table}[b!]
            \caption{Total fraction of UVOT-accessible GRBs detected after stacking the signal between elements in a nano-satellite constellation. The number in brackets indicates the detection fraction for high redshift ($z > 5$) bursts.}
            \centering
            \begin{tabular}{c|cc}
                \hline\hline
                & \multicolumn{2}{c}{TeleCommand Speed} \\
                \# Sats & Nominal & Instant \\
                \hline
                1  & 72.5\% (44.1\%) & 74.8\%  (47.2\%) \\
                2  & 79.7\% (52.5\%) & 82.9\% (58.7\%) \\
                3  & 82.6\% (57.4\%) & 85.4\% (63.1\%) \\
                5  & 86.0\% (63.3\%) & 88.1\% (68.2\%) \\
                10 & 89.6\% (70.4\%) & 91.0\% (74.6\%) \\
                \hline\hline
            \end{tabular}

            \label{tab:Total Multihopper Stacking}
        \end{table}
        
        % \begin{table}[t!]
        %     \caption{Total fraction of high redshift UVOT-accessible GRBs detected after stacking the signal between elements in a nano-satellite constellation.}
        %     \centering
        %     \begin{tabular}{c|cc}
        %         \hline\hline
        %         & \multicolumn{2}{c}{\textbf{TeleCommand Speed}} \\
        %         \textbf{\# Sats} & Nominal & Instant \\
        %         \hline
        %         1  & 44.1\% & 47.2\% \\
        %         2  & 52.5\% & 58.7\% \\
        %         3  & 57.4\% & 63.1\% \\
        %         5  & 63.3\% & 68.2\% \\
        %         10 & 70.4\% & 74.6\% \\
        %         \hline\hline
        %     \end{tabular}

        %     \label{tab:Total Multihopper Stacking}
        % \end{table}

        \begin{figure*}[hbt!]
            \centering
            \includegraphics[width=0.45\textwidth]{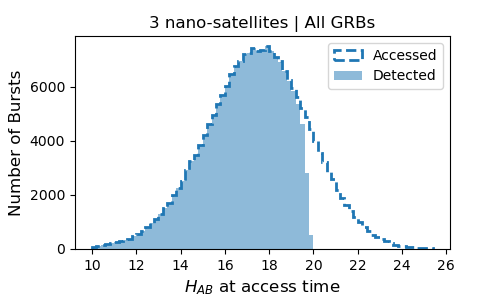}
            \includegraphics[width=0.45\textwidth]{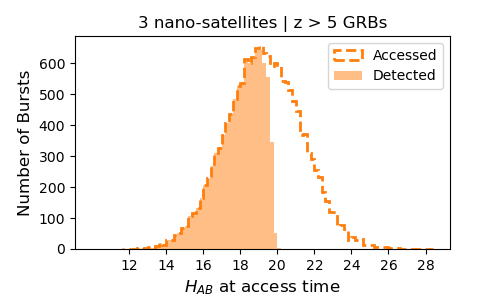}
            \caption{Histograms demonstrating the H-band afterglow magnitude of GRB afterglows at the time it was first accessed between a group of 3 near-infrared nano-satellites. The left plot depicts the full sample of $\sim 10^5$ afterglows, while the right plot shows only those events with $z > 5$. The dashed line indicates all afterglows generated in the simulation, and the shading indicates the GRBs which were detected (after stacking) by 3 nano-satellites using the $[1, 5, 5, ...]$ strategy with nominal TeleCommand latency.}
            \label{fig:3 skyhopper burst histograms}
        \end{figure*}
        
        The benefit to a larger constellation in the case of zero uplink latency comes from combining the signal from each satellite in the constellation. Combining the signal from 3 satellites allows the constellation to detect a total of $\sim 83\%$ of GRBs after signal stacking, whereas a 10-satellite constellation can detect $\sim 90\%$ of UVOT-observable Swift GRB triggers after stacking (Table \ref{tab:Total Multihopper Stacking}). The ability to combine signal between satellites becomes more valuable in the identification of high redshift events: a constellation of 3 satellites can detect $\sim 60\%$ of high redshift GRBs, and a constellation of 10 can detect as many as $70\%$ of high redshift bursts after combining the signal from the full constellation.
        
        In terms of cost effectiveness launching additional satellites may not be the best option, as a smaller number of satellites capable of instantaneous uplink capability can match or exceed the performance of larger constellations operating with nominal TeleCommand delay. Before stacking, a constellation of 2 satellites communicating instantly can detect $81.7\%$ of all bursts whereas a constellation of 10 satellites operating with nominal TeleCommand delay can only detect $82.7\%$. The larger constellations gain the benefits from combining signal between satellites, but even in this respect instantaneous communication is highly beneficial, as 5 satellites communicating instantly can achieve the same total detection fraction as 10 satellites with nominal communications delay (Table \ref{tab:Total Multihopper Stacking}).
        
        % The most economically efficient solution for rapid-response afterglow observations is likely to launch two satellites with a zero latency TeleCommand solution. This configuration is able to almost match the performance of larger constellations, achieving approximately the same number of detections within 100 minutes and only making $\sim 10\%$ fewer detections at very early times.
        
        Given the lack of a substantial advantage to larger constellations, we conclude that a group of 2-3 satellites (including a third one helps with redundancy/down-time of instruments) is the largest number of rapid-response near-infrared telescopes one should launch to observe GRB afterglows. Given this conclusion, we generate Figure \ref{fig:3 skyhopper burst histograms} in order to understand the performance increase gained by launching 3 satellites as compared to a single satellite. We find that a 3 satellite constellation can reliably detect every burst with a magnitude on access of $H_{AB} < 19$ (two magnitudes deeper than with a single satellite - Figure \ref{fig:skyhopper burst histograms}), when stacking the signal between satellites. Simultaneously, the mean magnitude of GRB afterglows at the time of access is $\sim 0.5$ magnitudes brighter in the case of 3 satellites compared to a single satellite, allowing the constellation to detect a greater fraction of the GRB population below the limiting magnitude of $H_{AB} = 20$ (the limiting magnitude to detection only increases as high as $H_{AB} \sim 20.6$ for a 10 satellite constellation).

        % \begin{figure}[t!]
        %     \centering
        %     \includegraphics[width = \columnwidth]{figures/3 SkyHopper Event Overview.png}
        %     \caption{Overview of every high redshift GRB observed by the near-infrared nano-satellite over 2 years of simulated observations. The x-axis shows the time that the near-infrared nano-satellite accessed the GRB, and the y-axis represents the $H$-band AB magnitude of the afterglow at the time of access. Circles indicate detections (using the $[1, 5, 5, ...]$ strategy with nominal TeleCommand latency), and crosses indicate non-detections.}
        %     \label{fig:3 SkyHopper Detection Overview}
        % \end{figure}

\section{Conclusion}
\label{sec:Conclusion}

    We present the results from a simplified mission simulation testing the ability of a rapid-response near-infrared nano-satellite (based on the SkyHopper mission concept) to perform follow-up observations on Swift GRB triggers. The simulation developed for this work combines an orbital line-of-sight modelling framework with Monte Carlo methods to randomly generate GRB events and average an infrared nano-satellite's performance over many simulated observing runs. We find that a near-infrared nano-satellite is able to not only expand the existing catalogue of high redshift GRBs, but also to make a significant and unique contribution to the study of early-time near-infrared GRB afterglows. Using the simulation framework outlined in this work we demonstrate the following:
    
    \begin{itemize}
    
        \item A near-infrared nano-satellite (using the optimal observing strategy) in a 550km polar Sun-synchronous orbit is capable of performing a successful follow-up afterglow detection on 72.5\% $\pm$ 3.1\% of UVOT-observable Swift GRB triggers within 2 hours of the initial burst, which corresponds to a GRB detection rate of $\sim 55 \pm 2$ GRBs yr$^{-1}$. Furthermore $\sim 30\%$ of Swift GRB triggers can be detected within $t = 10$ minutes of the GRB prompt emission, corresponding $\sim 20$ GRBs yr$^{-1}$ detected in this under-sampled region of GRB afterglow parameter space.
        
        \item A near-infrared nano-satellite can detect $\sim 1-3$ high redshift ($z > 5$) GRBs yr$^{-1}$. This predicted performance represents a substantial advance for the field, as at the time of writing only 23 such GRBs have been discovered over the last $\sim 24$ years of observations. 
        
        \item The optimal fixed observing strategy for early-time near-infrared afterglow observations is to first perform a 1 minute exposure of the target before transitioning to performing 5 minute exposures of the target. This strategy maximises the overall probability of detecting a GRB afterglow while also achieving a very high chance of a timely ($< 11$ minutes after GRB trigger) detection of the target. This strategy performs better than a strategy which attempts to take long exposures of the target given the variability in the observing window duration afforded from low Earth orbit.
        
        \item Launching a small constellation (2-3) of similar nano-satellites into orbit equally spaced around the same orbital plane is expected to double the early-time probability of detecting a GRB, and to increase the overall probability of detecting a GRB (after combining the signal from each satellite) from $\sim 72\%$ for a single satellite to $\sim 83\%$ for a  constellation of 3. Similarly, a 3-satellite constellation is expected to detect of $\sim 58\%$ of high redshift GRBs, compared to $\sim 44\%$ for a single satellite.
        %from $\sim 72\%$ for a single satellite to $\sim 83\%$ for a constellation of 3 (after combining the signal from each instrument). The same is true for high redshift bursts, where a 3-satellite constellation can detect $\sim 58\%$ of high-z GRBs compared to $\sim 44\%$ for a single satellite. 
        Launching more than 3 nano-satellites into orbit may not represent the best return on investment compared to reducing TeleCommand latency or increasing the point-source sensitivity of the instrument (i.e. increasing telescope aperture). In the case of a zero-latency TeleCommand solution, a constellation of 3 equally spaced satellites can equal the performance of a 10 satellite constellation with respect to timely GRB detections, though 10 satellites yield $\sim 10\%$ more GRB detections overall (i.e., $\sim 91\%$ probability of afterglow detection) after stacking the observations from every satellite.
    \end{itemize}
    
    Overall, we find that using a near-infrared nano-satellite for rapid-response GRB afterglow observations is not only a viable application for GRB science, but would be able to contribute substantially to the field via the identification of high redshift GRBs in a way that is not easily achievable from the ground. While a cost/benefit analysis is outside the scope of this work, the estimated order-of-magnitude cost of a SkyHopper-like nano-satellite mission is in the range of AU~\$10 million, making it competitive against the network of multiple 2m-class facilities on the ground that would be required to yield a comparable efficiency in GRB afterglow imaging follow-up. 
    
    Collecting a large database of high redshift GRBs is critical for astrophysical progress. These GRBs are cutting-edge opportunities to characterise the star formation history of the Universe back to the epoch of reionisation \citep{Trenti_2012}. They would enable the measurement of the chemical composition of intergalactic gas in the early Universe \citep{Cucchiara_2016} and the escape fraction of ionising radiation from galaxies \citep{Chen_2007}, which is one of the most challenging, yet fundamental, astronomical measurements. GRBs have distinct advantages as cosmological probes over quasars, as the latter carve out large ionised bubbles within their local environments \citep{Chornock_2013}, and the future of the field is bright with the James Webb Space Telescope and 30-m class facilities coming online with spectroscopic follow-up capabilities. 

% PASA uses footnotes, not endnotes. \endnote in this template will behave like \footnote; and \printendnotes will not output anything.
% \printendnotes

\section{Acknowledgements}
MTh. acknowledges support from an Australian Government Research Training Program (RTP) Scholarship. This research is supported in part by the Australian Research Council Centre of Excellence for All Sky Astrophysics in 3 Dimensions (ASTRO 3D), through project number CE170100013.

\bibliography{references}

\begin{thebibliography}{}
\expandafter\ifx\csname natexlab\endcsname\relax\def\natexlab#1{#1}\fi

\bibitem[{{Berger} {et~al.}(2005){Berger}, {Penprase}, {Fox}, {Kulkarni},
  {Hill}, {Schaefer}, \& {Reed}}]{Berger_2005}
{Berger}, E., {Penprase}, B.~E., {Fox}, D.~B., {et~al.} 2005, astro ph 0512280,
  arXiv:astro–ph/0512280

\bibitem[{{Chen} {et~al.}(2007){Chen}, {Prochaska}, \& {Gnedin}}]{Chen_2007}
{Chen}, H.-W., {Prochaska}, J.~X., \& {Gnedin}, N.~Y. 2007, \apjl, 667, L125

\bibitem[{{Chornock} {et~al.}(2013){Chornock}, {Berger}, {Fox}, {Lunnan},
  {Drout}, {Fong}, {Laskar}, \& {Roth}}]{Chornock_2013}
{Chornock}, R., {Berger}, E., {Fox}, D.~B., {et~al.} 2013, \apj, 774, 26

\bibitem[{{Cucchiara} {et~al.}(2016){Cucchiara}, {Totani}, \&
  {Tanvir}}]{Cucchiara_2016}
{Cucchiara}, A., {Totani}, T., \& {Tanvir}, N. 2016, \ssr, 202, 143

\bibitem[{Cucchiara {et~al.}(2011)Cucchiara, Levan, Fox, Tanvir, Ukwatta,
  Berger, Krühler, Yoldaş, Wu, Toma, \& et~al.}]{Cucchiara_2011}
Cucchiara, A., Levan, A.~J., Fox, D.~B., {et~al.} 2011, ApJ, 736, 7

\bibitem[{{Dressel}(2021)}]{Dressel_2021}
{Dressel}, L. 2021, Wide Field Camera 3 Instrument Handbook, Version 13.0

\bibitem[{Evans {et~al.}(2009)Evans, Beardmore, Page, Osborne, O'Brien,
  Willingale, Starling, Burrows, Godet, Vetere, Racusin, Goad, Wiersema,
  Angelini, Capalbi, Chincarini, Gehrels, Kennea, Margutti, Morris, Mountford,
  Pagani, Perri, Romano, \& Tanvir}]{Evans_2009}
Evans, P.~A., Beardmore, A.~P., Page, K.~L., {et~al.} 2009, Monthly Notices of
  the Royal Astronomical Society, 397, 1177

\bibitem[{Fiore {et~al.}(2020)Fiore, Burderi, Lavagna, Bertacin, Evangelista,
  Campana, Fuschino, Lunghi, Monge, Negri, \& et~al.}]{Fiore_2020}
Fiore, F., Burderi, L., Lavagna, M., {et~al.} 2020, Space Telescopes and
  Instrumentation 2020: Ultraviolet to Gamma Ray, doi:\url{10.1117/12.2560680}

\bibitem[{{Fox} {et~al.}(2008){Fox}, {Ledoux}, {Vreeswijk}, {Smette}, \&
  {Jaunsen}}]{Fox_2008}
{Fox}, A.~J., {Ledoux}, C., {Vreeswijk}, P.~M., {Smette}, A., \& {Jaunsen},
  A.~O. 2008, A\&A, 491, 189–207

\bibitem[{Gehrels {et~al.}(2009)Gehrels, Ramirez–Ruiz, \& Fox}]{Gehrels_2009}
Gehrels, N., Ramirez–Ruiz, E., \& Fox, D. 2009, ARA\&A, 47, 567–617

\bibitem[{{Gehrels} {et~al.}(2004){Gehrels}, {Chincarini}, {Giommi}, {Mason},
  {Nousek}, {Wells}, {White}, {Barthelmy}, {Burrows}, {Cominsky}, {Hurley},
  {Marshall}, {M{\'e}sz{\'a}ros}, {Roming}, {Angelini}, {Barbier}, {Belloni},
  {Campana}, {Caraveo}, {Chester}, {Citterio}, {Cline}, {Cropper}, {Cummings},
  {Dean}, {Feigelson}, {Fenimore}, {Frail}, {Fruchter}, {Garmire}, {Gendreau},
  {Ghisellini}, {Greiner}, {Hill}, {Hunsberger}, {Krimm}, {Kulkarni}, {Kumar},
  {Lebrun}, {Lloyd-Ronning}, {Markwardt}, {Mattson}, {Mushotzky}, {Norris},
  {Osborne}, {Paczynski}, {Palmer}, {Park}, {Parsons}, {Paul}, {Rees},
  {Reynolds}, {Rhoads}, {Sasseen}, {Schaefer}, {Short}, {Smale}, {Smith},
  {Stella}, {Tagliaferri}, {Takahashi}, {Tashiro}, {Townsley}, {Tueller},
  {Turner}, {Vietri}, {Voges}, {Ward}, {Willingale}, {Zerbi}, \&
  {Zhang}}]{Gehrels_2004}
{Gehrels}, N., {Chincarini}, G., {Giommi}, P., {et~al.} 2004, \apj, 611, 1005

\bibitem[{{Gehrels} {et~al.}(2005){Gehrels}, {Sarazin}, {O'Brien}, {Zhang},
  {Barbier}, {Barthelmy}, {Blustin}, {Burrows}, {Cannizzo}, {Cummings}, {Goad},
  {Holland}, {Hurkett}, {Kennea}, {Levan}, {Markwardt}, {Mason}, {Meszaros},
  {Page}, {Palmer}, {Rol}, {Sakamoto}, {Willingale}, {Angelini}, {Beardmore},
  {Boyd}, {Breeveld}, {Campana}, {Chester}, {Chincarini}, {Cominsky},
  {Cusumano}, {de Pasquale}, {Fenimore}, {Giommi}, {Gronwall}, {Grupe}, {Hill},
  {Hinshaw}, {Hjorth}, {Hullinger}, {Hurley}, {Klose}, {Kobayashi},
  {Kouveliotou}, {Krimm}, {Mangano}, {Marshall}, {McGowan}, {Moretti},
  {Mushotzky}, {Nakazawa}, {Norris}, {Nousek}, {Osborne}, {Page}, {Parsons},
  {Patel}, {Perri}, {Poole}, {Romano}, {Roming}, {Rosen}, {Sato}, {Schady},
  {Smale}, {Sollerman}, {Starling}, {Still}, {Suzuki}, {Tagliaferri},
  {Takahashi}, {Tashiro}, {Tueller}, {Wells}, {White}, \&
  {Wijers}}]{Gehrels_2005}
{Gehrels}, N., {Sarazin}, C.~L., {O'Brien}, P.~T., {et~al.} 2005, Nature, 437,
  851–854

\bibitem[{{Gordon} {et~al.}(2003){Gordon}, {Clayton}, {Misselt}, {Landolt}, \&
  {Wolff}}]{Gordon_2003}
{Gordon}, K.~D., {Clayton}, G.~C., {Misselt}, K.~A., {Landolt}, A.~U., \&
  {Wolff}, M.~J. 2003, \apj, 594, 279

\bibitem[{{Green} \& {Olszewski}(2020)}]{Green_2020}
{Green}, J.~D., \& {Olszewski}, H. 2020, Space Telescope WFC Instrument Science
  Report, 3

\bibitem[{{Greiner} {et~al.}(2008){Greiner}, {Bornemann}, {Clemens}, {Deuter},
  {Hasinger}, {Honsberg}, {Huber}, {Huber}, {Krauss}, {Kr{\"u}hler},
  {K{\"u}pc{\"u} Yolda{\textcommabelow s}}, {Mayer–Hasselwand er}, {Mican},
  {Primak}, {Schrey}, {Steiner}, {Szokoly}, {Th{\"o}ne}, {Yolda{\textcommabelow
  s}}, {Klose}, {Laux}, \& {Winkler}}]{Greiner_2008}
{Greiner}, J., {Bornemann}, W., {Clemens}, C., {et~al.} 2008, PASP, 120, 405

\bibitem[{{Greiner} {et~al.}(2009){Greiner}, {Kr{\"u}hler}, {Fynbo}, {Rossi},
  {Schwarz}, {Klose}, {Savaglio}, {Tanvir}, {McBreen}, {Totani}, {Zhang}, {Wu},
  {Watson}, {Barthelmy}, {Beardmore}, {Ferrero}, {Gehrels}, {Kann}, {Kawai},
  {Yolda{\c{s}}}, {M{\'e}sz{\'a}ros}, {Milvang-Jensen}, {Oates}, {Pierini},
  {Schady}, {Toma}, {Vreeswijk}, {Yolda{\c{s}}}, {Zhang}, {Afonso}, {Aoki},
  {Burrows}, {Clemens}, {Filgas}, {Haiman}, {Hartmann}, {Hasinger}, {Hjorth},
  {Jehin}, {Levan}, {Liang}, {Malesani}, {Pyo}, {Schulze}, {Szokoly}, {Terada},
  \& {Wiersema}}]{Greiner_2009}
{Greiner}, J., {Kr{\"u}hler}, T., {Fynbo}, J.~P.~U., {et~al.} 2009, \apj, 693,
  1610

\bibitem[{{Greiner} {et~al.}(2011){Greiner}, {Kr\"uhler, T.}, {Klose, S.},
  {Afonso, P.}, {Clemens, C.}, {Filgas, R.}, {Hartmann, D. H.}, {K\"upc\"u
  Yoldas, A.}, {Nardini, M.}, {, F. Olivares E.}, {Rau, A.}, {Rossi, A.},
  {Schady, P.}, \& {Updike, A.}}]{Greiner_2011}
{Greiner}, J., {Kr\"uhler, T.}, {Klose, S.}, {et~al.} 2011, AAP, 526, A30

\bibitem[{{Hartoog} {et~al.}(2015){Hartoog}, {Malesani}, {Fynbo}, {Goto},
  {Kr{\"u}hler}, {Vreeswijk}, {De Cia}, {Xu}, {M{\o}ller}, {Covino}, {D'Elia},
  {Flores}, {Goldoni}, {Hjorth}, {Jakobsson}, {Krogager}, {Kaper}, {Ledoux},
  {Levan}, {Milvang-Jensen}, {Sollerman}, {Sparre}, {Tagliaferri}, {Tanvir},
  {de Ugarte Postigo}, {Vergani}, {Wiersema}, {Datson}, {Salinas}, {Mikkelsen},
  \& {Aghanim}}]{Hartoog_2015}
{Hartoog}, O.~E., {Malesani}, D., {Fynbo}, J.~P.~U., {et~al.} 2015, \aap, 580,
  A139

\bibitem[{Kann {et~al.}(2006)Kann, Klose, \& Zeh}]{Kann_2006}
Kann, D.~A., Klose, S., \& Zeh, A. 2006, The Astrophysical Journal, 641,
  993–1009

\bibitem[{{Kawai} {et~al.}(2006){Kawai}, {Kosugi}, {Aoki}, {Yamada}, {Totani},
  {Ohta}, {Iye}, {Hattori}, {Aoki}, {Furusawa}, {Hurley}, {Kawabata},
  {Kobayashi}, {Komiyama}, {Mizumoto}, {Nomoto}, {Noumaru}, {Ogasawara},
  {Sato}, {Sekiguchi}, {Shirasaki}, {Suzuki}, {Takata}, {Tamagawa}, {Terada},
  {Watanabe}, {Yatsu}, \& {Yoshida}}]{Kawai_2006}
{Kawai}, N., {Kosugi}, G., {Aoki}, K., {et~al.} 2006, \nat, 440, 184

\bibitem[{{Klose} {et~al.}(2004){Klose}, {Palazzi}, {Masetti}, {Stecklum},
  {Greiner}, {Hartmann}, \& {Schmid}}]{Klose_2004}
{Klose}, S., {Palazzi}, E., {Masetti}, N., {et~al.} 2004, \aap, 420, 899

\bibitem[{{Klose} {et~al.}(2000){Klose}, {Stecklum}, {Masetti}, {Pian},
  {Palazzi}, {Henden}, {Hartmann}, {Fischer}, {Gorosabel},
  {S{\'a}nchez-Fern{\'a}ndez}, {Butler}, {Ott}, {Hippler}, {Kasper}, {Weiss},
  {Castro-Tirado}, {Greiner}, {Bartolini}, {Guarnieri}, {Piccioni}, {Benetti},
  {Ghinassi}, {Magazz{\'u}}, {Hurley}, {Cline}, {Trombka}, {McClanahan},
  {Starr}, {Goldsten}, {Gold}, {Mazets}, {Golenetskii}, {Noeske}, {Papaderos},
  {Vreeswijk}, {Tanvir}, {Oscoz}, {Mu{\~n}oz}, \& {Castro
  Cer{\'o}n}}]{Klose_2000}
{Klose}, S., {Stecklum}, B., {Masetti}, N., {et~al.} 2000, \apj, 545, 271

\bibitem[{{Knapp} {et~al.}(2020){Knapp}, {Seager}, {Demory}, {Krishnamurthy},
  {Smith}, {Pong}, {Bailey}, {Donner}, {Pasquale}, {Campuzano}, {Smith}, {Luu},
  {Babuscia}, {Bocchino}, {Loveland}, {Colley}, {Gedenk}, {Kulkarni}, {Hughes},
  {White}, {Krajewski}, \& {Fesq}}]{knapp2020}
{Knapp}, M., {Seager}, S., {Demory}, B.-O., {et~al.} 2020, \aj, 160, 23

\bibitem[{{Kr\"uhler} {et~al.}(2011){Kr\"uhler}, {Schady, P.}, {Greiner, J.},
  {Afonso, P.}, {Bottacini, E.}, {Clemens, C.}, {Filgas, R.}, {Klose, S.},
  {Koch, T. S.}, {K\"upc\"u–Yoldas, A.}, {Oates, S. R.}, {, F. Olivares E.},
  {Page, M. J.}, {McBreen, S.}, {Nardini, M.}, {Nicuesa Guelbenzu, A.}, {Rau,
  A.}, {Roming, P. W. A.}, {Rossi, A.}, {Updike, A.}, \& {Yoldas,
  A.}}]{Kruhler_2011}
{Kr\"uhler}, T., {Schady, P.}, {Greiner, J.}, {et~al.} 2011, A\&A, 526, A153

\bibitem[{Kumar \& Zhang(2015)}]{Kumar_2015}
Kumar, P., \& Zhang, B. 2015, Physics Reports, 561, 1–109

\bibitem[{Lamb \& Reichart(2000)}]{Lamb_2000}
Lamb, D.~Q., \& Reichart, D.~E. 2000, ApJ, 536, 1

\bibitem[{{Lidz} {et~al.}(2021){Lidz}, {Chang}, {Mas-Ribas}, \&
  {Sun}}]{Lidz_2021}
{Lidz}, A., {Chang}, T.-C., {Mas-Ribas}, L., \& {Sun}, G. 2021, \apj, 917, 58

\bibitem[{{McGuire}(2016)}]{McGuire_2016}
{McGuire}, J.~T.~W. 2016, in Eighth Huntsville Gamma-Ray Burst Symposium, Vol.
  1962, 4021

\bibitem[{{McQuinn} {et~al.}(2008){McQuinn}, {Lidz}, {Zaldarriaga},
  {Hernquist}, \& {Dutta}}]{McQuinn_2008}
{McQuinn}, M., {Lidz}, A., {Zaldarriaga}, M., {Hernquist}, L., \& {Dutta}, S.
  2008, MNRAS, 388, 1101–1110

\bibitem[{Mearns \& Trenti(2018)}]{Mearns_2018}
Mearns, R., \& Trenti, M. 2018, arXiv e-print, arXiv:1808.06746,
  arXiv:1808.06746

\bibitem[{{Mesinger} \& {Furlanetto}(2008)}]{Mesinger_2008}
{Mesinger}, A., \& {Furlanetto}, S.~R. 2008, MNRAS, 385, 1348–1358

\bibitem[{Miralda‐Escude(1998)}]{Miralda_Escude_1998}
Miralda‐Escude, J. 1998, ApJ, 501, 15–22

\bibitem[{{Oates} {et~al.}(2009){Oates}, {Page}, {Schady}, {de Pasquale},
  {Koch}, {Breeveld}, {Brown}, {Chester}, {Holland}, {Hoversten}, {Kuin},
  {Marshall}, {Roming}, {Still}, {vanden Berk}, {Zane}, \&
  {Nousek}}]{Oates_2009}
{Oates}, S.~R., {Page}, M.~J., {Schady}, P., {et~al.} 2009, MNRAS, 395,
  490–503

\bibitem[{Oates {et~al.}(2015)Oates, Racusin, De~Pasquale, Page, Castro-Tirado,
  Gorosabel, Smith, Breeveld, \& Kuin}]{Oates_2015}
Oates, S.~R., Racusin, J.~L., De~Pasquale, M., {et~al.} 2015, Monthly Notices
  of the Royal Astronomical Society, 453, 4121

\bibitem[{{Perley} {et~al.}(2013){Perley}, {Levan}, {Tanvir}, {Cenko}, {Bloom},
  {Hjorth}, {Kr{\"u}hler}, {Filippenko}, {Fruchter}, {Fynbo}, {Jakobsson},
  {Kalirai}, {Milvang-Jensen}, {Morgan}, {Prochaska}, \&
  {Silverman}}]{Perley_2013}
{Perley}, D.~A., {Levan}, A.~J., {Tanvir}, N.~R., {et~al.} 2013, \apj, 778, 128

\bibitem[{{Perley} {et~al.}(2016){Perley}, {Kr{\"u}hler}, {Schulze}, {de Ugarte
  Postigo}, {Hjorth}, {Berger}, {Cenko}, {Chary}, {Cucchiara}, {Ellis}, {Fong},
  {Fynbo}, {Gorosabel}, {Greiner}, {Jakobsson}, {Kim}, {Laskar}, {Levan},
  {Micha{\l}owski}, {Milvang-Jensen}, {Tanvir}, {Th{\"o}ne}, \&
  {Wiersema}}]{Perley_2016}
{Perley}, D.~A., {Kr{\"u}hler}, T., {Schulze}, S., {et~al.} 2016, \apj, 817, 7

\bibitem[{{Prochaska} {et~al.}(2008){Prochaska}, {Dessauges–Zavadsky},
  {Ramirez–Ruiz}, \& {Chen}}]{Prochaska_2008}
{Prochaska}, J.~X., {Dessauges–Zavadsky}, M., {Ramirez–Ruiz}, E., \&
  {Chen}, H. 2008, ApJ, 685, 344–353

\bibitem[{{Racusin} {et~al.}(2017){Racusin}, {Perkins}, {Briggs}, {de Nolfo},
  {Krizmanic}, {Caputo}, {McEnery}, {Shawhan}, {Morris}, {Connaughton},
  {Kocevski}, {Wilson-Hodge}, {Hui}, {Mitchell}, \& {McBreen}}]{Racusin_2017}
{Racusin}, J., {Perkins}, J.~S., {Briggs}, M.~S., {et~al.} 2017, arXiv
  e-prints, arXiv:1708.09292

\bibitem[{Rhodes {et~al.}(2020)Rhodes, van der Horst, Fender, Monageng,
  Anderson, Antoniadis, Bietenholz, Böttcher, Bright, Green, \&
  et~al.}]{Rhodes_2020}
Rhodes, L., van der Horst, A.~J., Fender, R., {et~al.} 2020, Monthly Notices
  of the Royal Astronomical Society, 496, 3326–3335

\bibitem[{{Robertson} \& {Ellis}(2012)}]{Robertson_2012}
{Robertson}, B.~E., \& {Ellis}, R.~S. 2012, \apj, 744, 95

\bibitem[{Rueda {et~al.}(2018)Rueda, Ruffini, Wang, Aimuratov, de~Almeida,
  Bianco, Chen, Lobato, Maia, Primorac, \& et~al.}]{Rueda_2018}
Rueda, J., Ruffini, R., Wang, Y., {et~al.} 2018, JCAP, 2018, 006

\bibitem[{{Salvaterra} {et~al.}(2009){Salvaterra}, {Della Valle}, {Campana},
  {Chincarini}, {Covino}, {D'Avanzo}, {Fern{\'a}ndez-Soto}, {Guidorzi},
  {Mannucci}, {Margutti}, {Th{\"o}ne}, {Antonelli}, {Barthelmy}, {de Pasquale},
  {D'Elia}, {Fiore}, {Fugazza}, {Hunt}, {Maiorano}, {Marinoni}, {Marshall},
  {Molinari}, {Nousek}, {Pian}, {Racusin}, {Stella}, {Amati}, {Andreuzzi},
  {Cusumano}, {Fenimore}, {Ferrero}, {Giommi}, {Guetta}, {Holland}, {Hurley},
  {Israel}, {Mao}, {Markwardt}, {Masetti}, {Pagani}, {Palazzi}, {Palmer},
  {Piranomonte}, {Tagliaferri}, \& {Testa}}]{Salvaterra_2009}
{Salvaterra}, R., {Della Valle}, M., {Campana}, S., {et~al.} 2009, \nat, 461,
  1258

\bibitem[{Sari {et~al.}(1998)Sari, Piran, \& Narayan}]{Sari_1998}
Sari, R., Piran, T., \& Narayan, R. 1998, ApJ, 497, L17–L20

\bibitem[{Schady(2017)}]{Schady_2017}
Schady, P. 2017, arXiv e-prints, arXiv:1707.05214

\bibitem[{{Skrutskie} {et~al.}(2006){Skrutskie}, {Cutri}, {Stiening},
  {Weinberg}, {Schneider}, {Carpenter}, {Beichman}, {Capps}, {Chester},
  {Elias}, {Huchra}, {Liebert}, {Lonsdale}, {Monet}, {Price}, {Seitzer},
  {Jarrett}, {Kirkpatrick}, {Gizis}, {Howard}, {Evans}, {Fowler}, {Fullmer},
  {Hurt}, {Light}, {Kopan}, {Marsh}, {McCallon}, {Tam}, {Van Dyk}, \&
  {Wheelock}}]{Skrutskie_2006}
{Skrutskie}, M.~F., {Cutri}, R.~M., {Stiening}, R., {et~al.} 2006, \aj, 131,
  1163

\bibitem[{Stanway {et~al.}(2008)Stanway, Bremer, \& Lehnert}]{Stanway_2008}
Stanway, E.~R., Bremer, M.~N., \& Lehnert, M.~D. 2008, MNRAS, 385, 493–510

\bibitem[{{Tagliaferri} {et~al.}(2005){Tagliaferri}, {Antonelli}, {Chincarini},
  {Fern{\'a}ndez-Soto}, {Malesani}, {Della Valle}, {D'Avanzo}, {Grazian},
  {Testa}, {Campana}, {Covino}, {Fiore}, {Stella}, {Castro-Tirado},
  {Gorosabel}, {Burrows}, {Capalbi}, {Cusumano}, {Conciatore}, {D'Elia},
  {Filliatre}, {Fugazza}, {Gehrels}, {Goldoni}, {Guetta}, {Guziy}, {Held},
  {Hurley}, {Israel}, {Jel{\'\i}nek}, {Lazzati}, {L{\'o}pez-Echarri},
  {Melandri}, {Mirabel}, {Moles}, {Moretti}, {Mason}, {Nousek}, {Osborne},
  {Pellizza}, {Perna}, {Piranomonte}, {Piro}, {de Ugarte Postigo}, \&
  {Romano}}]{Tagliaferri_2005}
{Tagliaferri}, G., {Antonelli}, L.~A., {Chincarini}, G., {et~al.} 2005, \aap,
  443, L1

\bibitem[{{Tanvir} {et~al.}(2009){Tanvir}, {Fox}, {Levan}, {Berger},
  {Wiersema}, {Fynbo}, {Cucchiara}, {Kr{\"u}hler}, {Gehrels}, {Bloom},
  {Greiner}, {Evans}, {Rol}, {Olivares}, {Hjorth}, {Jakobsson}, {Farihi},
  {Willingale}, {Starling}, {Cenko}, {Perley}, {Maund}, {Duke}, {Wijers},
  {Adamson}, {Allan}, {Bremer}, {Burrows}, {Castro-Tirado}, {Cavanagh}, {de
  Ugarte Postigo}, {Dopita}, {Fatkhullin}, {Fruchter}, {Foley}, {Gorosabel},
  {Kennea}, {Kerr}, {Klose}, {Krimm}, {Komarova}, {Kulkarni}, {Moskvitin},
  {Mundell}, {Naylor}, {Page}, {Penprase}, {Perri}, {Podsiadlowski}, {Roth},
  {Rutledge}, {Sakamoto}, {Schady}, {Schmidt}, {Soderberg}, {Sollerman},
  {Stephens}, {Stratta}, {Ukwatta}, {Watson}, {Westra}, {Wold}, \&
  {Wolf}}]{Tanvir_2009}
{Tanvir}, N.~R., {Fox}, D.~B., {Levan}, A.~J., {et~al.} 2009, \nat, 461, 1254

\bibitem[{{Tanvir} {et~al.}(2012){Tanvir}, {Levan}, {Fruchter}, {Fynbo},
  {Hjorth}, {Wiersema}, {Bremer}, {Rhoads}, {Jakobsson}, {O'Brien}, {Stanway},
  {Bersier}, {Natarajan}, {Greiner}, {Watson}, {Castro-Tirado}, {Wijers},
  {Starling}, {Misra}, {Graham}, \& {Kouveliotou}}]{Tanvir_2012}
{Tanvir}, N.~R., {Levan}, A.~J., {Fruchter}, A.~S., {et~al.} 2012, \apj, 754,
  46

\bibitem[{Trenti {et~al.}(2015)Trenti, Perna, \& Jimenez}]{Trenti_2015}
Trenti, M., Perna, R., \& Jimenez, R. 2015, ApJ, 802, 103

\bibitem[{{Trenti} {et~al.}(2012){Trenti}, {Perna}, {Levesque}, {Shull}, \&
  {Stocke}}]{Trenti_2012}
{Trenti}, M., {Perna}, R., {Levesque}, E.~M., {Shull}, J.~M., \& {Stocke},
  J.~T. 2012, \apjl, 749, L38

\bibitem[{{Trenti} {et~al.}(2013){Trenti}, {Perna}, \&
  {Tacchella}}]{Trenti_2013}
{Trenti}, M., {Perna}, R., \& {Tacchella}, S. 2013, \apjl, 773, L22

\bibitem[{{Troja}(2020)}]{Troja_2020}
{Troja}, E. 2020, The Neil Gehrels Swift Observatory Technical Handbook Version
  17.0

\bibitem[{{van Eerten} {et~al.}(2010){van Eerten}, {Zhang}, \&
  {MacFadyen}}]{Eerten_2010}
{van Eerten}, H., {Zhang}, W., \& {MacFadyen}, A. 2010, \apj, 722, 235

\bibitem[{{von Kienlin} {et~al.}(2020){von Kienlin}, {Meegan}, {Paciesas},
  {Bhat}, {Bissaldi}, {Briggs}, {Burns}, {Cleveland}, {Gibby}, {Giles},
  {Goldstein}, {Hamburg}, {Hui}, {Kocevski}, {Mailyan}, {Malacaria},
  {Poolakkil}, {Preece}, {Roberts}, {Veres}, \&
  {Wilson-Hodge}}]{von_Kienlin_2020}
{von Kienlin}, A., {Meegan}, C.~A., {Paciesas}, W.~S., {et~al.} 2020, \apj,
  893, 46

\bibitem[{{Vreeswijk} {et~al.}(2007){Vreeswijk}, {Ledoux}, {Smette}, {Ellison},
  {Jaunsen}, {Andersen}, {Fruchter}, {Fynbo}, {Hjorth}, {Kaufer}, {M{\o}ller},
  {Petitjean}, {Savaglio}, \& {Wijers}}]{Vreeswijk_2007}
{Vreeswijk}, P.~M., {Ledoux}, C., {Smette}, A., {et~al.} 2007, \aap, 468, 83

\bibitem[{{Wanderman} \& {Piran}(2010)}]{Wanderman_2010}
{Wanderman}, D., \& {Piran}, T. 2010, MNRAS, 406, 1944

\bibitem[{Wang {et~al.}(2021{\natexlab{a}})Wang, Yang, Fan, Hennawi, Barth,
  Banados, Bian, Boutsia, Connor, Davies, \& et~al.}]{Wang_2021}
Wang, F., Yang, J., Fan, X., {et~al.} 2021{\natexlab{a}}, ApJ, 907, L1

\bibitem[{Wang {et~al.}(2013)Wang, Liang, Li, Lu, Wei, \& Zhang}]{Wang_2013}
Wang, X.-G., Liang, E.-W., Li, L., {et~al.} 2013, The Astrophysical Journal,
  774, 132

\bibitem[{Wang {et~al.}(2021{\natexlab{b}})Wang, Zheng, Xiao, Yang, Liu, Yang,
  Zou, Zhang, Zeng, Xiong, Feng, Song, Wen, Xu, Chen, Ni, Zhang, Wu, Cai, Cang,
  Deng, Gao, Kong, Huang, Li, Li, Li, Liang, Lin, Liu, Long, Lu, Luo, Ma, Meng,
  Peng, Qiao, Song, Tian, Wang, Wang, Wang, Xu, Yang, Yin, Zeng, Zeng, Zhang,
  Zhang, Zhang, \& Zhang}]{Wang_2021b}
Wang, X.~I., Zheng, X., Xiao, S., {et~al.} 2021{\natexlab{b}}, The
  Astrophysical Journal, 922, 237

\bibitem[{Wei {et~al.}(2016)Wei, Cordier, Antier, Antilogus, Atteia, Bajat,
  Basa, Beckmann, Bernardini, Boissier, Bouchet, Burwitz, Claret, Dai, Daigne,
  Deng, Dornic, Feng, Foglizzo, Gao, Gehrels, Godet, Goldwurm, Gonzalez,
  Gosset, Götz, Gouiffes, Grise, Gros, Guilet, Han, Huang, Huang, Jouret,
  Klotz, Marle, Lachaud, Floch, Lee, Leroy, Li, Li, Li, Liang, Lyu, Mercier,
  Migliori, Mochkovitch, O'Brien, Osborne, Paul, Perinati, Petitjean, Piron,
  Qiu, Rau, Rodriguez, Schanne, Tanvir, Vangioni, Vergani, Wang, Wang, Wang,
  Wang, Watson, Webb, Wei, Willingale, Wu, Wu, Xin, Xu, Yu, Yu, Yu, Zhang,
  Zhang, Zhang, \& Zhou}]{Wei_2016}
Wei, J., Cordier, B., Antier, S., {et~al.} 2016, arXiv e-prints,
  arXiv:1610.06892

\bibitem[{{Wen} {et~al.}(2019){Wen}, {Long}, {Zheng}, {An}, {Cai}, {Cang},
  {Che}, {Chen}, {Chen}, {Chen}, {Chen}, {Cheng}, {Deng}, {Deng}, {Ding}, {Du},
  {Duan}, {Gan}, {Gao}, {Gao}, {Han}, {Han}, {He}, {He}, {Hou}, {Hu}, {Hu},
  {Huang}, {Huang}, {Huang}, {Jia}, {Jiang}, {Jin}, {Li}, {Li}, {Li}, {Liang},
  {Liang}, {Lin}, {Liu}, {Liu}, {Liu}, {Liu}, {Liu}, {Liu}, {Lu}, {Lu}, {Lu},
  {Luo}, {Ma}, {Ma}, {Mao}, {Mo}, {Nie}, {Qu}, {Shan}, {Shi}, {Song}, {Sun},
  {Tan}, {Tang}, {Tao}, {Wang}, {Wang}, {Wang}, {Wu}, {Wu}, {Xia}, {Xiao},
  {Xie}, {Xu}, {Xu}, {Xu}, {Yan}, {Yan}, {Yang}, {Yang}, {Yang}, {Yang},
  {Yang}, {Yao}, {Yu}, {Yu}, {Zhang}, {Zhang}, {Zhang}, {Zhang}, {Zhang},
  {Zhang}, {Zhang}, {Zhao}, {Zhao}, {Zheng}, {Zhou}, {Zhu}, {Zou}, {An}, {Cai},
  {Chen}, {Dai}, {Fan}, {Feng}, {Feng}, {Gao}, {Huang}, {Kang}, {Li}, {Li},
  {Liang}, {Lin}, {Lin}, {Liu}, {Liu}, {Liu}, {Liu}, {Lu}, {Mao}, {Shen},
  {Shu}, {Su}, {Sun}, {Tam}, {Tang}, {Tian}, {Wang}, {Wang}, {Wang}, {Wang},
  {Wu}, {Wu}, {Xiong}, {Xu}, {Yu}, {Yu}, {Yu}, {Zeng}, {Zeng}, {Zhang},
  {Zhang}, {Zhao}, {Zhou}, \& {Zhu}}]{Wen_2019}
{Wen}, J., {Long}, X., {Zheng}, X., {et~al.} 2019, Experimental Astronomy, 48,
  77

\bibitem[{Werner {et~al.}(2018)Werner, Řípa, Pál, Ohno, Tarcai, Torigoe,
  Tanaka, Uchida, Mészáros, Galgóczi, Fukazawa, Mizuno, Takahashi, Nakazawa,
  Várhegyi, Enoto, Odaka, Ichinohe, Frei, \& Kiss}]{Werner_2018}
Werner, N., Řípa, J., Pál, A., {et~al.} 2018, in Space Telescopes and
  Instrumentation 2018: Ultraviolet to Gamma Ray, ed. J.-W.~A. den Herder,
  S.~Nikzad, \& K.~Nakazawa, Vol. 10699, International Society for Optics and
  Photonics (SPIE), 672 -- 686

\bibitem[{{Woosley} \& {Bloom}(2006)}]{Woosley_2006}
{Woosley}, S.~E., \& {Bloom}, J.~S. 2006, \araa, 44, 507

\end{thebibliography}

% \appendix

% \input{example-appendices}
\end{document}